\pdfoutput=1

\RequirePackage{fix-cm}

\documentclass[twocolumn]{svjour3}

\smartqed

\usepackage{graphicx}
\usepackage{amssymb,amsmath}
\usepackage{txfonts}
\usepackage[numbers]{natbib}
\usepackage{hyperref}

\journalname{Nonlinear Dynamics}

\begin{document}

\title{Fractal basin boundaries and escape dynamics in a multiwell potential}

\author{Euaggelos E. Zotos}

\institute{Department of Physics, School of Science, \\
Aristotle University of Thessaloniki, \\
GR-541 24, Thessaloniki, Greece \\
Corresponding author's email: {evzotos@physics.auth.gr}}

\date{Received: 13 January 2016 / Accepted: 31 March 2016 / Published online: 21 April 2016}

\titlerunning{Fractal basin boundaries and escape dynamics in a multiwell potential}

\authorrunning{Euaggelos E. Zotos}

\maketitle

\begin{abstract}

The escape dynamics in a two-dimensional multiwell potential is explored. A thorough numerical investigation is conducted in several types of two-dimensional planes and also in a three-dimensional subspace of the entire four-dimensional phase space in order to distinguish between non-escaping (ordered and chaotic) and escaping orbits. The determination of the location of the basins of escape towards the different escape channels and their correlations with the corresponding escape time of the orbits is undoubtedly an issue of paramount importance. It was found that in all examined cases regions of non-escaping motion coexist with several basins of escape. Furthermore, we monitor how the percentages of all types of orbits evolve when the total orbital energy varies. The larger escape periods have been measured for orbits with initial conditions in the fractal basin boundaries, while the lowest escape rates belong to orbits with initial conditions inside the basins of escape. The Newton-Raphson basins of attraction of the equilibrium points of the system have also been determined. We hope that our numerical analysis will be useful for a further understanding of the escape mechanism of orbits in open Hamiltonian systems with two degrees of freedom.


\end{abstract}

\section{Introduction}
\label{intro}

Particles moving in escaping orbits in Hamiltonian systems is, without any doubt, one of the most interesting topics in nonlinear dynamics (e.g., \cite{C90,CK92,CKK93,GSP07,GSRL11,STN02}). Hamiltonian systems with escapes are also known as open or leaking Hamiltonian systems, in which there is a finite energy of escape. When the value of the energy is higher than the energy of escape, the equipotential surfaces open and escape channels emerge through which the test particle can escape to infinity. It should emphasize that if a test particle has energy larger than the escape value, this does not necessarily mean that it will certainly escape from the system and even if escape does occur, the time required for the escape to occur may be very long compared with the natural crossing time. The literature is replete with research studies on the field of leaking Hamiltonian systems (e.g., \cite{BBS09,BC96,EP14,KSCD99,NH01,STUL95,T89,Z14a,Z14b,Z15b}).

The problem of escaping orbits in open Hamiltonian systems is however less explored than the related problem of chaotic scattering. The viewpoint of chaos theory has been used in order to investigate and interpret the phenomenon of chaotic scattering (e.g., \cite{BTS96,BST98,BOG89,JLS99,JMS95,JT91,SASL06,SSL07}). At this point, we would like to emphasize that all the above-mentioned references on the issues of open Hamiltonian systems and chaotic scattering are exemplary rather than exhaustive.

In leaking Hamiltonian systems an issue of paramount importance is the determination of the basins of escape, similar to basins of attraction in dissipative systems or even the Newton-Raphson fractal structures. An escape basin is defined as a local set of initial conditions of orbits for which the test particles escape through a certain exit in the equipotential surface for energies above the escape value. Basins of escape have been studied in many earlier papers (e.g., \cite{BGOB88,C02,KY91,PCOG96,SO00}). The reader can find more details regarding basins of escape in \cite{C02}, while the review \cite{Z14b} provides useful information about the escape properties of orbits in a multi-channel dynamical system of a two-dimensional perturbed harmonic oscillator. The boundaries between the escape basins may be fractal (e.g., \cite{AVS09,BGOB88,dML99}) or even respect the more restrictive Wada property (e.g., \cite{AVS01}), in the case where three or more escape channels coexist in the equipotential surface.

In two recent papers \cite{Z14b,Z15b} we investigated the escape dynamics in multi channel potentials composed of perturbed harmonic oscillators. In this paper we shall use the same computational methods in order to reveal the escape mechanism of orbits in the umbilical catastrophe potential $D_5$ which has two wells (e.g., \cite{BBI06,BCIK10}). The dynamics of transition between different equilibrium states, such as nuclear fission, chemical reactions and phase transitions, can be realistically modeled by Hamiltonian systems with multiwell potentials. Catastrophe theory analyses degenerate critical points of the potential function, which means that points where not just the first derivative, but one or more higher derivatives of the potential function are also zero. Umbilic catastrophes are examples of rank 2 catastrophes and there are three main categories: (i) the hyperbolic umbilic catastrophe, (ii) the elliptic umbilic catastrophe, and (iii) the parabolic umbilic catastrophe. The $D_5$ potential belongs to the third category.

The structure of the paper is as follows: In Section \ref{mod} we present in detail the properties of the Hamiltonian system. The next Section is devoted on the Newton-Raphson basins of attraction, while all the computational methods we used in order to explore the escape dynamics of the orbits are described in Section \ref{cometh}. In the following Section, we conduct a thorough and systematic numerical investigation revealing the escape mechanism of the $D_5$ potential. Our paper ends with Section \ref{disc} where the discussion of our research is given.

\section{Presentation of the model potential}
\label{mod}

We shall investigate the escape dynamics in a characteristic example of a two-dimensional (2D) multiwell potential which is the lower umbilical catastrophe $D_5$
\begin{equation}
V_{D_5}(x,y) = 2\alpha y^2 - x^2 + xy^2 + \frac{1}{4} x^4.
\label{pot}
\end{equation}
Without the loss of generality we can assume that $\alpha = 1$. However, it should be underlined that by fixing the value of $\alpha$ the conclusions drawn are specific of the considered system and they cannot be generalized.

The equations of motion governing the motion of a test particle with a unit mass $(m = 1)$ are
\begin{equation}
V_x = \ddot{x} = - \frac{\partial V_{D_5}}{\partial x}, \ \ \
V_y = \ddot{y} = - \frac{\partial V_{D_5}}{\partial y},
\label{eqmot}
\end{equation}
where, as usual, the dot indicates derivative with respect to the time. Furthermore, the variational equations needed for the computation of standard chaos indicators (the SALI\footnote{The SALI method was chosen over more classical dynamical indicators (e.g., the positive Lyapunov exponent, the fast fourier transform, etc) for distinguishing between order and chaos because it can automatically classify initial conditions of orbits using only the final numerical value of SALI at the end of the numerical integration. On the other hand, for all other classical methods we need to plot either the time-evolution of the indicator (e.g., the positive Lyapunov exponent) or the shape of the spectrum (e.g., the fast fourier transform) in order to determine the character of an orbit. Obviously, this is not possible when we have to classify large sets of initial conditions of orbits. In this case, the ideal solution is a ``one-number index" such as the SALI.} in our case, as better explained in the following section) are given by
\begin{eqnarray}
\dot{(\delta x)} &=& \delta \dot{x}, \ \ \ \dot{(\delta y)} = \delta \dot{y}, \nonumber \\
(\dot{\delta \dot{x}}) &=& -\frac{\partial^2 V_{D_5}}{\partial x^2}\delta x - \frac{\partial^2 V_{D_5}}{\partial x \partial y}\delta y, \nonumber \\
(\dot{\delta \dot{y}}) &=& -\frac{\partial^2 V_{D_5}}{\partial y \partial x}\delta x - \frac{\partial^2 V_{D_5}}{\partial y^2}\delta y.
\label{variac}
\end{eqnarray}

\begin{figure}[!t]
\begin{center}
\includegraphics[width=\hsize]{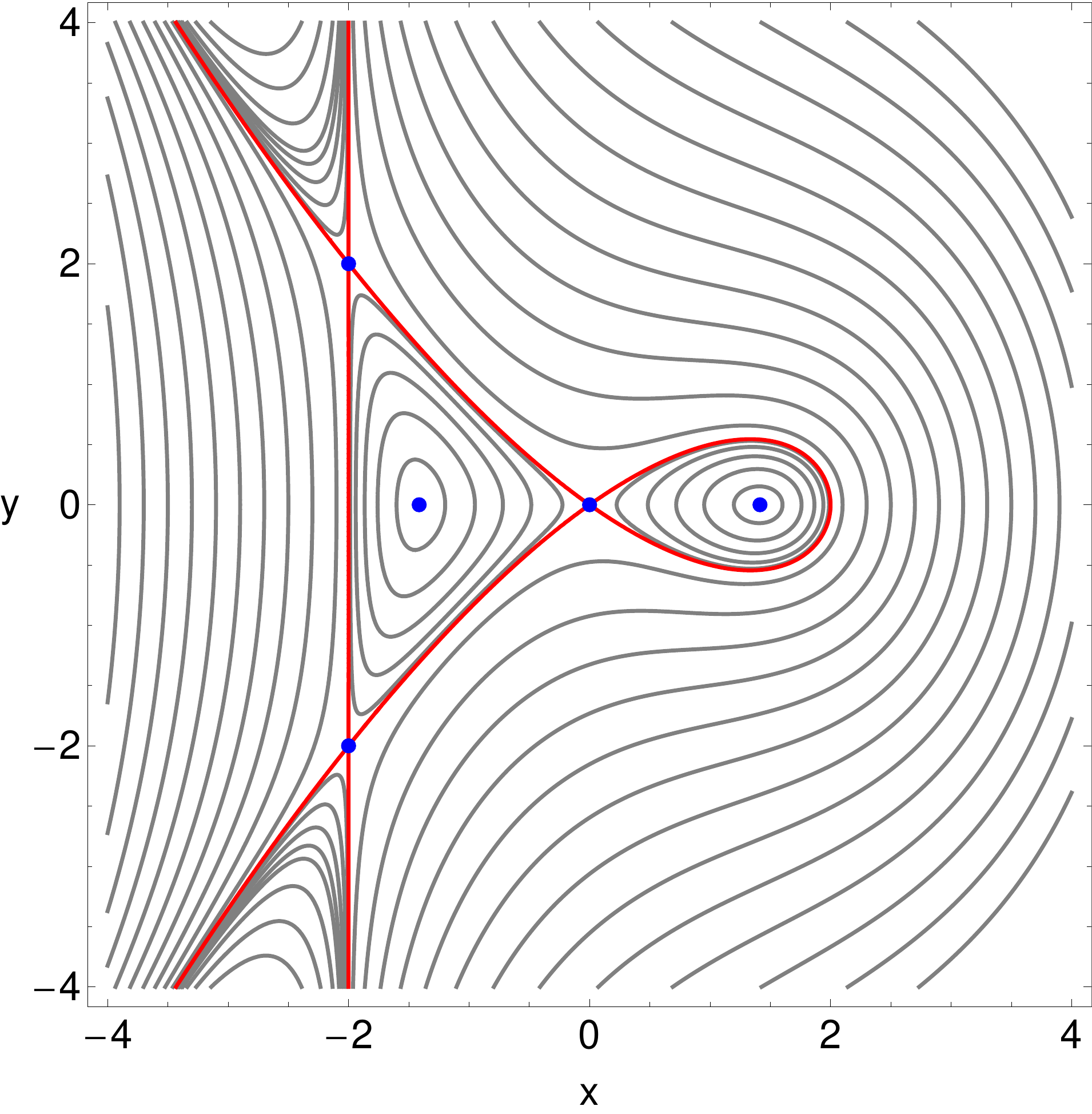}
\end{center}
\caption{The isoline contours of the effective potential $V_{D_5}$ in the configuration $(x,y)$ space. The isoline contour corresponding to the critical energy of escape $E_{esc}$ is shown in red. Included are the five equilibrium points the positions of which are represented by blue dots.}
\label{isopot}
\end{figure}

\begin{figure*}[!t]
\centering
\resizebox{\hsize}{!}{\includegraphics{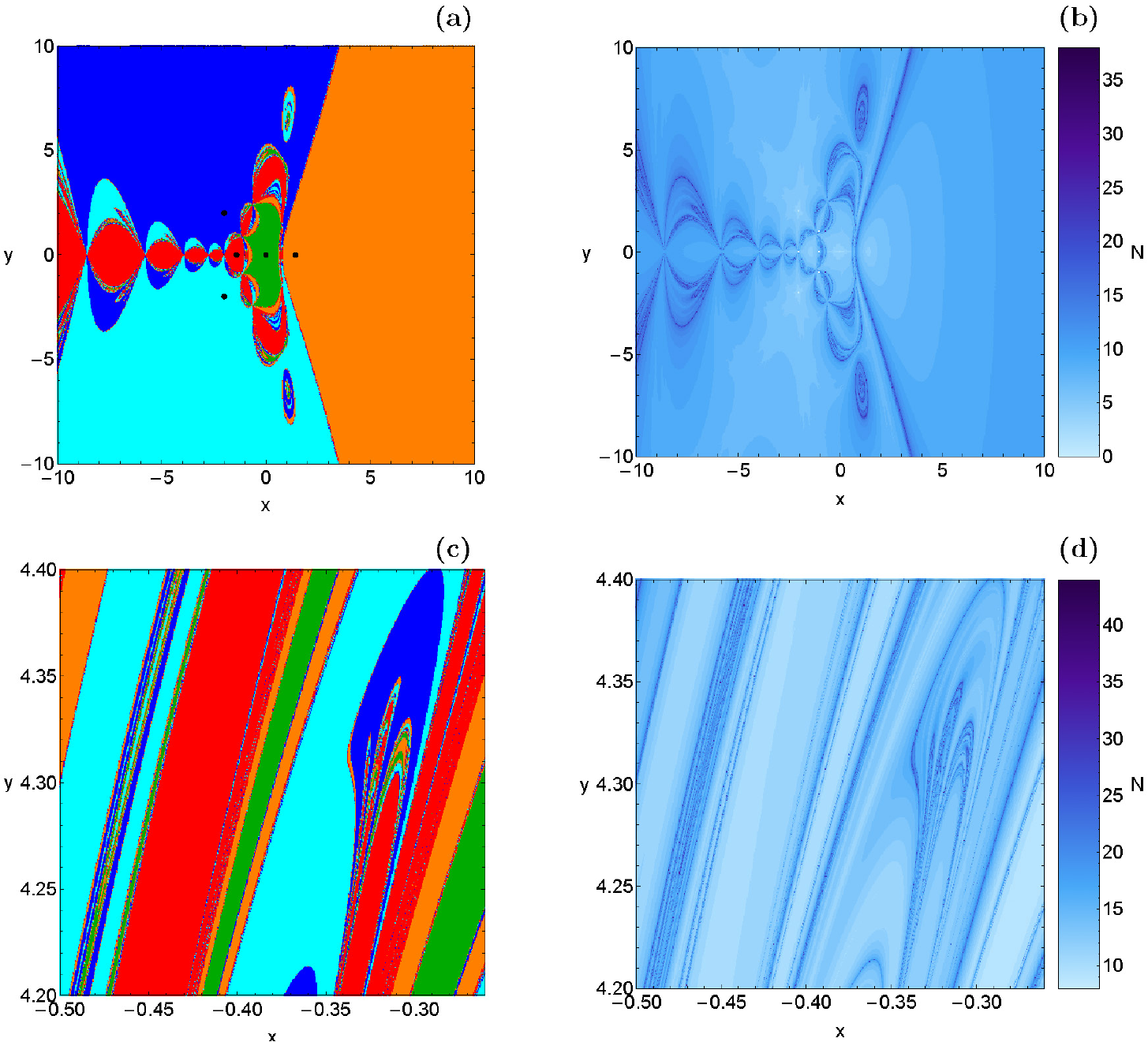}}
\caption{(a-upper left): The Newton-Raphson basins of attraction of the $D_5$ potential on the configuration $(x,y)$ plane. The positions of the five equilibrium points are indicated by black dots. (b-upper right): the corresponding number $(N)$ of required iterations. (c-lower right): a magnification of panel (a). (d-lower right): the corresponding number $(N)$ of required iterations.}
\label{nr}
\end{figure*}

The Hamiltonian to potential (\ref{pot}) reads
\begin{equation}
H(x,y,\dot{x},\dot{y}) = \frac{1}{2}\left(\dot{x}^2 + \dot{y}^2 \right) + V_{D_5}(x,y) = E,
\label{ham}
\end{equation}
where $\dot{x}$ and $\dot{y}$ are the velocities, while $E$ is the numerical value of the Hamiltonian, which is conserved. Thus, an orbit with a given value of energy is restricted in its motion to regions in which $E \leq V_{D_5}(x,y)$, while all other regions are energetically forbidden to the test particle.

The $D_5$ potential has five equilibrium points at which
\begin{equation}
\frac{\partial V_{D_5}}{\partial x} = \frac{\partial V_{D_5}}{\partial y} = 0.
\label{sys}
\end{equation}
Two local minima located at $(\pm \sqrt{2},0)$ and three saddles at $(0,0)$, and $(-2,\pm 2)$. Potential (\ref{pot}) has a finite energy of escape which is $E_{esc} = 0$. In Fig. \ref{isopot} we present the isoline contours of the effective potential in the configuration $(x,y)$ space. The isoline contour corresponding to the critical energy of escape $E_{esc}$ is shown in red, while the positions of the five equilibrium points are pinpointed by blue dots. Looking at Fig. \ref{isopot} we see that the central region of the potential is composed of two lobes. A transport channel between the two lobes is present only when $E > 0$.

\section{Newton-Raphson basins of attraction}
\label{bas}

In the previous section we found that the $D_5$ potential has five equilibrium points which are the solutions of the system of algebraic equations (\ref{sys}). We decided to use the multivariate Newton-Raphson method, a simple yet a very accurate computational tool, in order to determine to which of the five equilibrium points each initial point on the configuration $(x,y)$ plane leads to. The multivariate Newton-Raphson method takes the form
\begin{eqnarray}
x_{n+1} &=& x_n - \left( \frac{V_x V_{yy} - V_y V_{xy}}{V_{yy} V_{xx} - V^2_{xy}} \right)_{(x_n,y_n)}, \nonumber\\
y_{n+1} &=& y_n + \left( \frac{V_x V_{yx} - V_y V_{xx}}{V_{yy} V_{xx} - V^2_{xy}} \right)_{(x_n,y_n)},
\label{nrm}
\end{eqnarray}
where $x_n$, $y_n$ are the values of the $x$ and $y$ variables at the $n$-th step of the iterative process, while the subscripts of $V$ denote the corresponding partial derivatives. The reader can find more information regarding the derivation of Eqs. (\ref{nrm}) in the Appendix. The multivariate Newton-Raphson method has also been used to obtain the basins of attraction in other dynamical systems, such as the restricted three-body problem (e.g., \cite{KGK12}), or the four-body problem (e.g., \cite{KK14}).

The Newton-Raphson algorithm is activated when an initial condition $(x_0,y_0)$ on the configuration plane is given, while it stops when the positions of the equilibrium points are reached, with some predefined accuracy. All the initial conditions that lead to a specific equilibrium point, compose a basin of attraction or an attracting region. Here we would like to clarify that the Newton-Raphson basins of attraction should not be mistaken with the classical basins of attraction in dissipative systems. We observe that the iterative formulae (\ref{nrm}) include both the first and the second derivatives of the $D_5$ potential and therefore we may claim that the obtained numerical results directly reflect some of the basic qualitative characteristics of the Hamiltonian system. The major advantage of knowing the Newton-Raphson basins of attraction in a dynamical system is the fact that we can select the most favorable initial conditions, with respect to required computation time, when searching for an equilibrium point.

For obtaining the Newton-Raphson basins of attraction we worked as follows: First we defined a dense uniform grid of $1024 \times 1024$ initial conditions regularly distributed on the configuration space. The iterative process was terminated when an accuracy of $10^{-15}$ has been reached, while we classified all the $(x,y)$ initial conditions that lead to a particular solution (equilibrium point). At the same time, for each initial point, we recorded the number $(N)$ of iterations required to obtain the aforementioned accuracy. Logically, the required number of iterations for locating an equilibrium point strongly depends on the value of the predefined accuracy. In Fig. \ref{nr}a we present the Newton-Raphson basins of attractions for the $D_5$ potential, while in Fig. \ref{nr}b we provide the corresponding number $(N)$ of required iterations. We observe that several areas of the configuration plane is covered by broad well-defined basins of attractions, while there are also regions in which it is impossible to predict to which equilibrium point each initial condition leads to. In Fig. \ref{nr}c, where a magnification of a specific area on the $(x,y)$ plane is depicted, we can observe the fractal boundaries between the several basins of attraction.

\section{Computational methods}
\label{cometh}

In order to explore the escape dynamics of the $D_5$ multiwell potential we need to define sets of initial conditions of orbits. For this task we define for each value of the energy integral of motion (all tested energy levels are above the escape energy), dense uniform grids of $1024 \times 1024$ initial conditions regularly distributed in the area allowed by the value of the energy $E$. Our investigation takes place in several types of planes in order to obtain a spherical and a more complete view of the escape process of the $D_5$ potential.

\begin{figure}[!t]
\begin{center}
\includegraphics[width=\hsize]{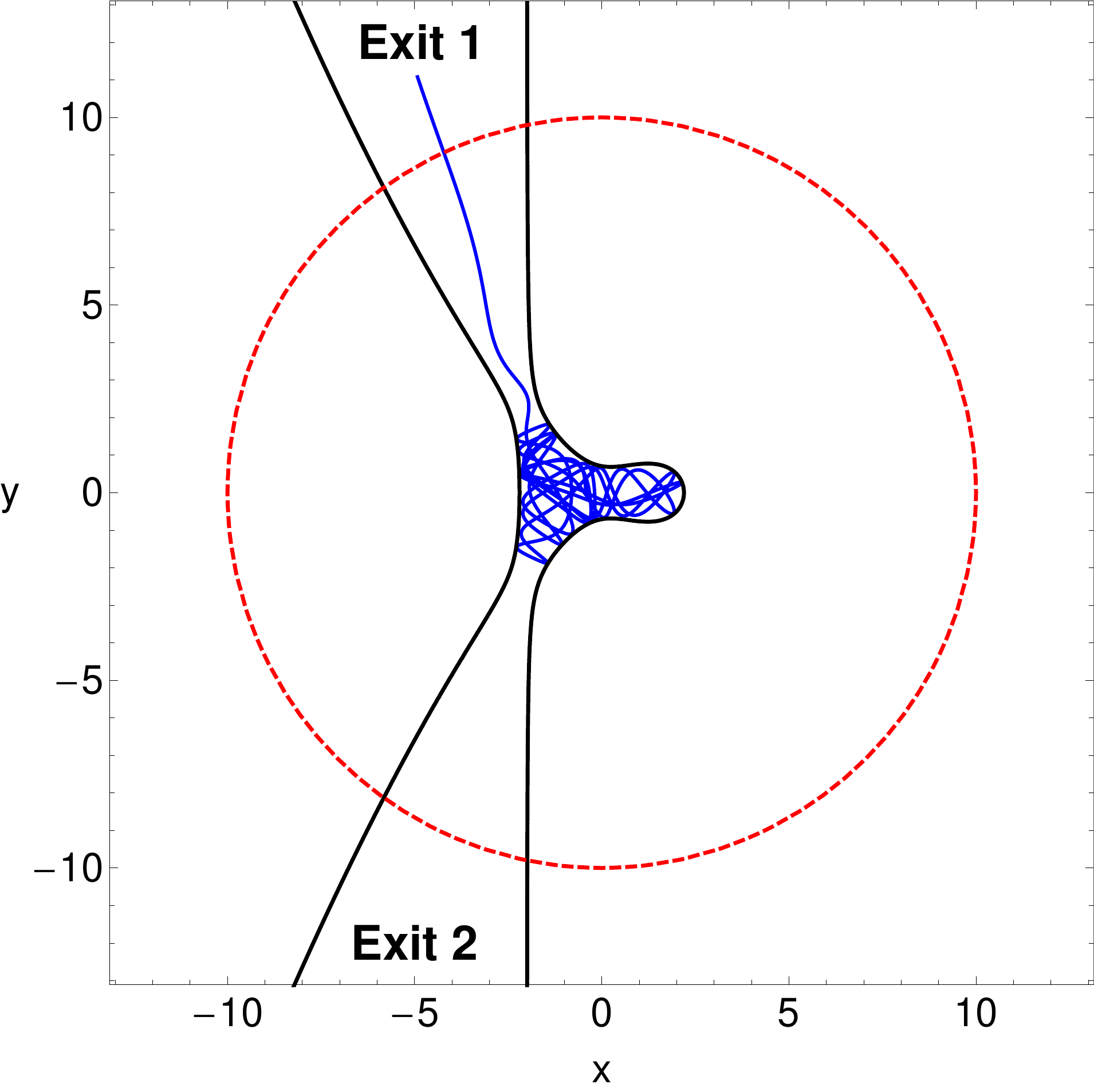}
\end{center}
\caption{The isoline contour (black) of the $D_5$ potential when $E = 1$. The orbit intersects the limiting circle (red) and escapes from exit channel 1.}
\label{orb}
\end{figure}

An issue of paramount importance is the determination of the position as well as the time at which an orbit escapes. When the value of the total orbital energy $E$ is smaller than the escape energy, the Zero Velocity Curves (ZVCs) are closed. On the other hand, when $E > E_{esc}$ the ZVCs are open and extend to infinity. An open ZVC consists of several branches forming channels through which an orbit can escape to infinity. At every opening there is a highly unstable periodic orbit close to the line of maximum potential \cite{C79} which is called a Lyapunov orbit \cite{L49}. Such an orbit reaches the ZVC, on both sides of the opening and returns along the same path thus, connecting two opposite branches of the ZVC. Usually the Lyapunov orbits are used to determine the escapes of orbits. In particular, an orbit is considered as an escaping one when it intersects one of the Lyapunov orbits with velocity pointing outwards. However their use has a disadvantage since it can be used only for orbits with initial conditions inside the central region of the potential. But what about orbits with initial conditions outside the Lyapunov orbits? Do all these orbits escape? Could some of these orbits move inside the central region? In order to give answers to these questions we shall apply the geometrical escape criterion used successfully in \cite{SS08}. According to this criterion an orbit is considered to escape when $x^2 + y^2 > R^2$, where $R = 10$. This allows us to correctly determine the escape of orbits with initial conditions inside a scattering region of a limit circle with radius $R = 10$. Fig. \ref{orb} shows how an orbit with initial conditions: $x_0 = -0.43, y_0 = \dot{x_0} = 0, \dot{y_0} > 0$ when $E = 1$ intersects the limiting circle and escapes from exit channel 1 after about 54 time units of numerical integration.

In Hamiltonian systems the configuration as well as the phase space is divided into the escaping and non-escaping (trapped) regions. Usually, the vast majority of the non-escaping space is occupied by initial conditions of regular orbits forming stability islands where a third adelphic integral of motion is present. In many systems however, trapped chaotic orbits have also been observed (e.g., \cite{Z15a}). Therefore, we decided to distinguish between regular non-escaping and trapped chaotic orbits. Over the years, several chaos indicators have been developed in order to determine the character of orbits. In our case, we chose to use the Smaller ALingment Index (SALI) method. The SALI \cite{S01} has been proved a very fast, reliable and effective tool. The value of the SALI indicates the character of an orbit. In particular, if SALI $> 10^{-4}$ the orbit if regular, while if SALI $< 10^{-8}$ the orbit is chaotic. When $10^{-4} \leq$ SALI $\leq 10^{-8}$ we have the case of a ``sticky orbit" and further numerical integration is needed so as the true nature of the orbit to be fully revealed. Sticky orbits initially behave as regular ones, while their true chaotic character is revealed only after a long time interval of numerical integration.

For the numerical integration we set a maximum time equal to $10^4$ time units. Our previous experience in this subject indicates that usually orbits need considerable less time to find one of the exits in the equipotential surface and eventually escape from the system (obviously, the numerical integration is effectively ended when an orbit passes through one of the escape channels and intersects the limit circle). Nevertheless, we decided to use such a vast integration time just to be sure that all orbits have enough time in order to escape. Here we should clarify that orbits which do not escape after a numerical integration of $10^4$ time units are considered as non-escaping or trapped.

A double precision Bulirsch-Stoer \verb!FORTRAN 77! algorithm (e.g., \cite{PTVF92}) was used in order to forward integrate the equations of motion (\ref{eqmot}) as well as the variational equations (\ref{variac}) for all the initial conditions of the orbits. Throughout all our computations, the energy integral of motion of Eq. (\ref{ham}) was conserved better than one part in $10^{-12}$, although for most orbits it was better than one part in $10^{-13}$. All graphics presented in this work have been created using version 10.3 of Mathematica$^{\circledR}$ \cite{W03}.

\section{Escape dynamics}
\label{numres}

Our main target in this section will be to distinguish between escaping and non-escaping orbits for values of energy larger than the escape energy where the ZVCs are open and two channels of escape are present. Furthermore, two important properties of the orbits will be investigated: (i) the directions or channels through which the test particles escape and (ii) the time-scale of the escapes (we shall also use the term escape period). In particular, we will examine these dynamical quantities for various values of the total orbital energy $E$.

Our initial numerical calculations indicate that apart from the escaping orbits there is also an amount of non-escaping orbits. In general terms, the majority of non-escaping regions corresponds to initial conditions of regular orbits, where a third integral of motion is present, restricting their accessible phase space and therefore hinders their escape. However, there are also chaotic orbits which do not escape within the predefined time interval and remain trapped for vast periods until they eventually escape to infinity. At this point, it should be emphasized and clarified that these trapped chaotic orbits cannot be considered, by no means, neither as sticky orbits nor as super sticky orbits with sticky periods larger than $10^4$ time units. Sticky orbits are those who behave regularly for long time periods before their true chaotic nature is fully revealed. In our case on the other hand, this type of orbits exhibit chaoticity very quickly as it takes no more than about 100 time units for the SALI to cross the threshold value (SALI $\ll 10^{-8}$), thus identifying beyond any doubt their chaotic character. Therefore, we decided to classify the initial conditions of orbits into three main categories: (i) orbits that escape through one of the two escape channels, (ii) non-escaping regular orbits and (iii) trapped chaotic orbits.

\begin{figure*}[!t]
\centering
\resizebox{\hsize}{!}{\includegraphics{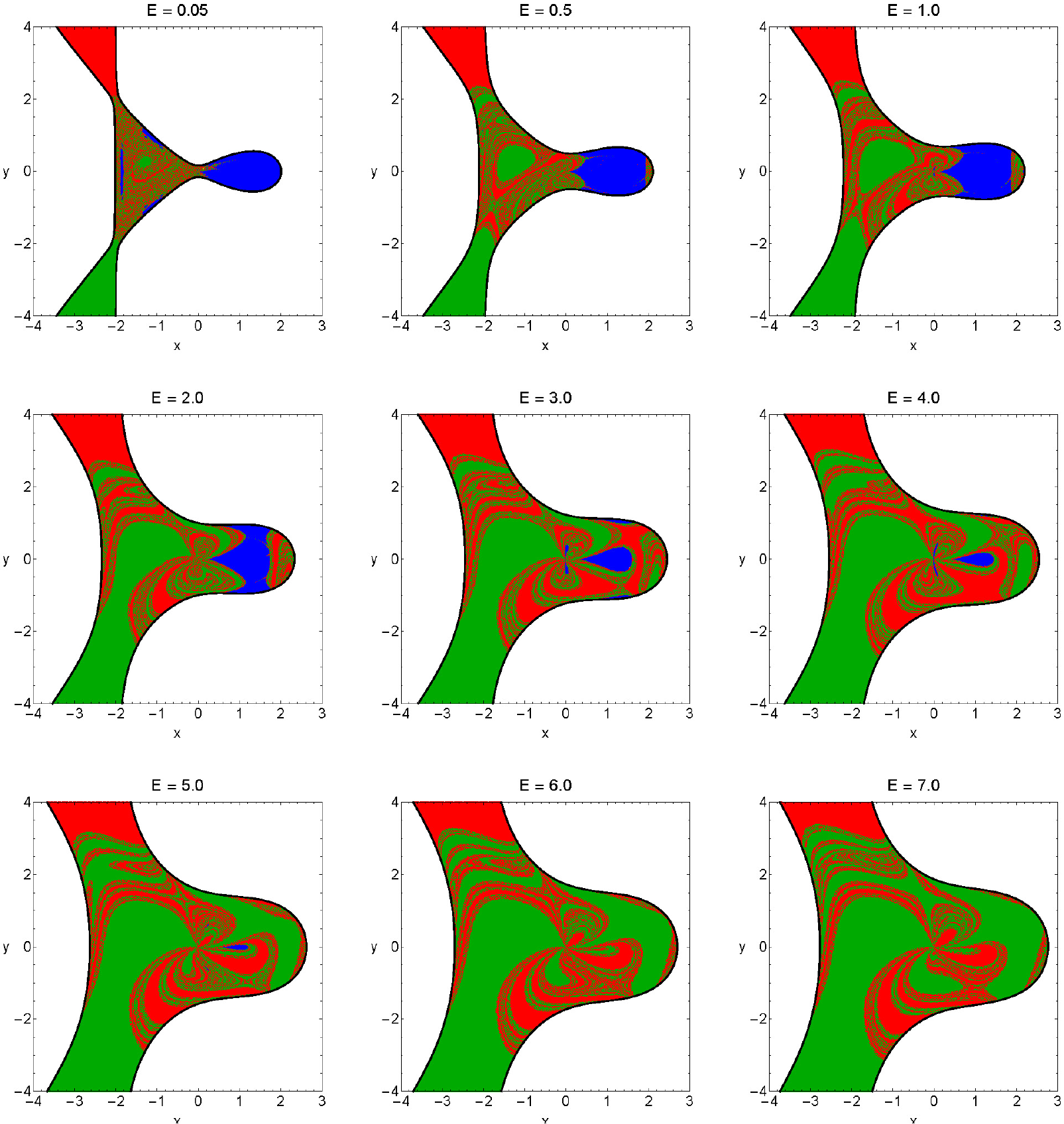}}
\caption{The orbital structure of the configuration $(x,y)$ plane for several values of the energy $E$. The color code is as follows: Escape through channel 1 (red); escape through channel 2 (green); non-escaping regular (blue); trapped chaotic (yellow).}
\label{xy}
\end{figure*}

\begin{figure*}[!t]
\centering
\resizebox{\hsize}{!}{\includegraphics{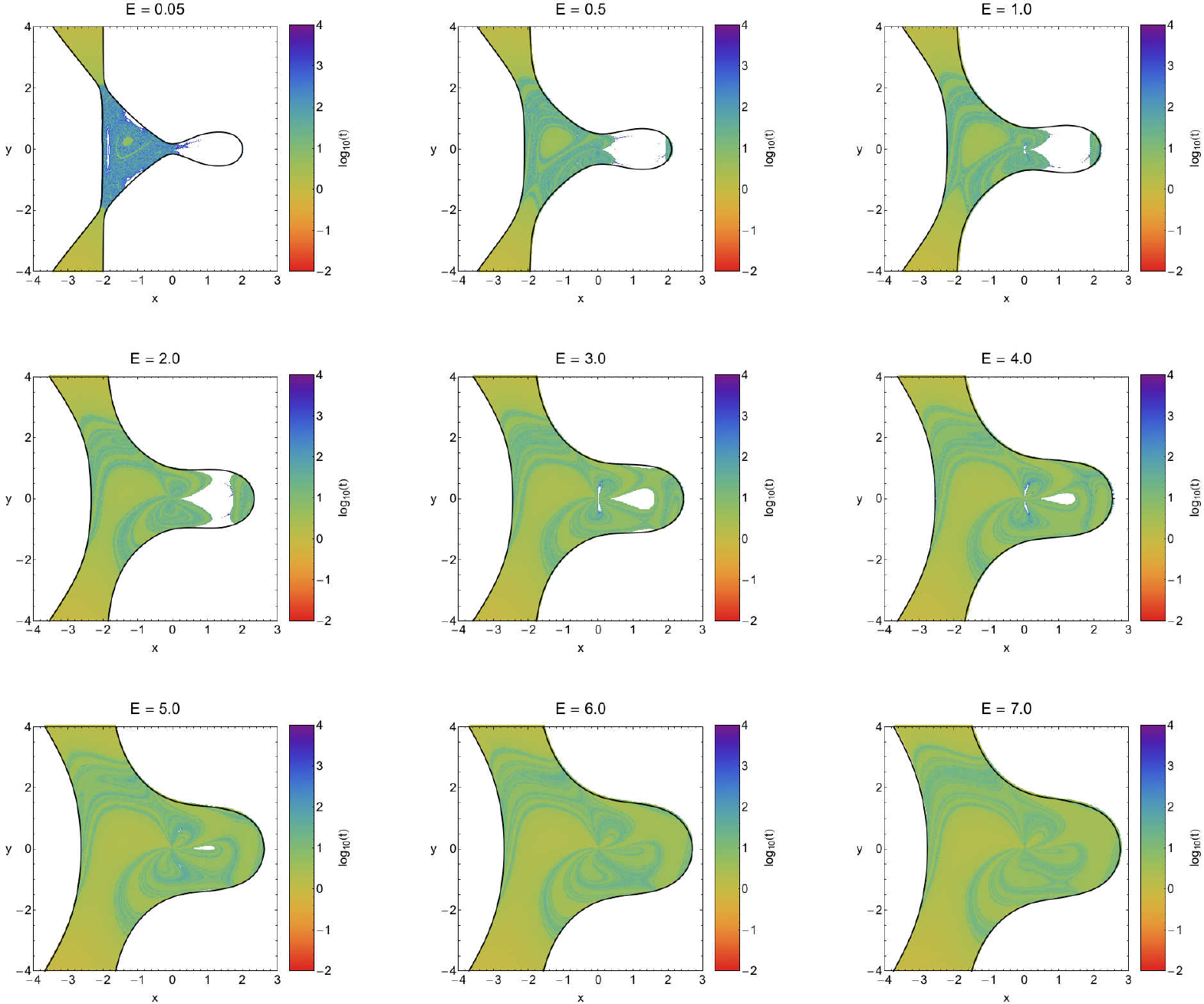}}
\caption{Distribution of the escape time $t_{\rm esc}$ of the orbits on the configuration $(x,y)$ plane. The bluer the color, the larger the escape time. Trapped chaotic and non-escaping regular orbits are shown in white.}
\label{xyt}
\end{figure*}

\subsection{Results for the configuration $(x,y)$ space}
\label{ss1}

Our exploration begins in the configuration $(x,y)$ space and in Fig. \ref{xy} we present the orbital structure of the $(x,y)$ plane for values of energy in the set $E$ = \{0.05, 0.5, 1, 2, 3, 4, 5, 6, 7\}. The sets of the initial conditions of the orbits are defined as follows: In polar coordinates $(r,\phi)$ the condition $\dot{r} = 0$ defines a two-dimensional surface of section, with two disjoint parts $\dot{\phi} < 0$ and $\dot{\phi} > 0$. Each of these two parts has a unique projection onto the configuration $(x,y)$ space. We chose to work on the $\dot{\phi} > 0$ part. The conditions $\dot{\phi} > 0$ and $\dot{r} = 0$ along with the existence of the integral of motion (\ref{ham}) suggest that the four initial conditions of orbits in cartesian coordinates are
\begin{eqnarray}
x &=& x_0, \nonumber\\
y &=& y_0, \nonumber\\
\dot{x_0} &=& - \frac{y_0}{r}\sqrt{2(E-V_{D_5}(x_0,y_0))}, \nonumber\\
\dot{y_0} &=& \frac{x_0}{r}\sqrt{2(E-V_{D_5}(x_0,y_0))},
\end{eqnarray}
where $r = \sqrt{x_0^2 + y_0^2}$.
Each initial condition is colored according to the escape channel through which the particular orbit escapes. The blue regions on the other hand, denote initial conditions where the test particles move in regular orbits and do not escape, while trapped chaotic orbits are indicated in yellow. The outermost solid line is the ZVC which is defined as $V_{D_5}(x,y) = E$.

It is seen that for $E = 0.05$, that is an energy level just above the energy of escape, inside the left lobe of the central region of the potential there is a highly sensitive dependence of the escape process on the initial conditions. Indeed a slight change in the initial conditions makes the test particle escape through the opposite channel, which is of course a classical indication of chaos. The right lobe on the other hand, is covered almost entirely by initial conditions of non-escaping regular orbits. As we proceed to higher energy levels two important phenomena take place: (i) the amount of non-escaping regular orbits is reduced and for $E \geq 6$ there is no indication of bounded motion or whatsoever, (ii) the fractal regions in the central region are also reduce and several basins of escape emerge. By the term basin of escape, we refer to a local set of initial conditions that corresponds to a certain escape channel. Here we would like to emphasize that when we state that an area is fractal we simply mean that it has a fractal-like geometry without conducting any specific calculations as in \cite{AVS09}. The fractality is strongly related with the unpredictability in the evolution of a Hamiltonian system. In our case, it can be interpreted that for high enough energy levels the test particles escape very fast from the scattering region and therefore, the predictability of the dynamical system increases. Looking at Fig. \ref{xy} we see that in all cases (energy levels) the boundaries between the escape basins are fractal. The existence of fractal basin boundaries is a very common phenomenon observed in leaking Hamiltonian systems (e.g., \cite{BGOB88,dML99,dMG02,LT11,STN02,ST03,TSPT04}). However with increasing energy the basin boundaries become more more smooth which means that the degree of fractality reduces.

The distribution of the escape time $t_{\rm esc}$ of orbits on the configuration $(x,y)$ space is given in the Fig. \ref{xyt}, where light reddish colors correspond to fast escaping orbits, dark blue, purple colors indicate large escape periods, while white color denote both trapped chaotic and non-escaping regular orbits. It is observed that for $h = 0.05$, that is a value of energy just above the escape energy, the escape periods of the majority of orbits with initial conditions in the central regions of the potential are huge corresponding to tens of thousands of time units. This however, is anticipated because in this case the width of the two escape channels is very small and therefore, the orbits should spend much time inside the ZVC until they find one of the two openings and eventually escape to infinity. As the value of the energy increases however, the escape channels become more and more wide leading to faster escaping orbits, which means that the escape period decreases rapidly. We found that the longest escape rates correspond to initial conditions near the vicinity of the fractal regions. On the other hand, the shortest escape periods have been measured for the regions without sensitive dependence on the initial conditions (basins of escape), that is, those far away from the fractal basin boundaries.

\begin{figure}[!t]
\begin{center}
\includegraphics[width=\hsize]{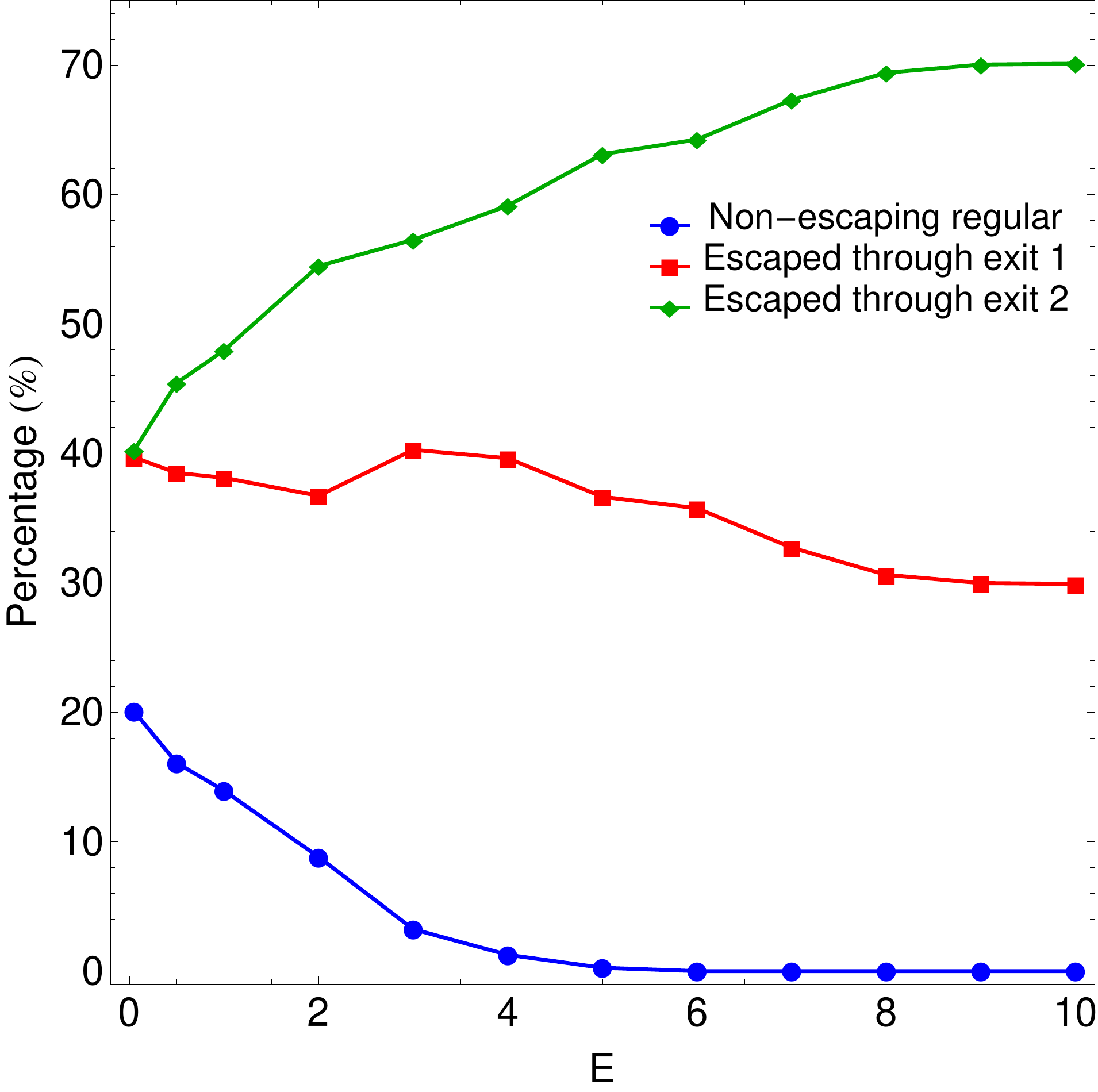}
\end{center}
\caption{Evolution of the percentages of escaping and non-escaping orbits with initial conditions on the configuration $(x,y)$ space when varying the total orbital energy $E$.}
\label{p1}
\end{figure}

It would be very informative to monitor the evolution of the percentages of all types of orbits as a function of the total orbital energy $E$. In the following Fig. \ref{p1} we present such a diagram. At this point we would like to point out that we decided not to include the percentage of trapped chaotic orbits because our computations indicate that always the rate of trapped chaotic orbits is extremely small (less than 0.1\%) and therefore, it does not contribute to the overall orbital structure of the Hamiltonian system. We observe that for $E = 0.05$ the escaping orbits share about 80\% of the configuration space. This is anticipated because at low energy levels the degree of fractality is high and therefore both channels are equiprobable. However as the value of the energy increases the percentages of escaping orbits start to diverge. In particular, the rate of escaping orbits through exit 1 is reduced, while on the other hand the amount of escaping orbits through exit 2 grows and at relatively high energy levels $(E > 8)$ it seems to saturates at about 70\%. The percentage of non-escaping regular orbits starts at 20\%, just above the energy of escape and then it drops as we proceed to higher energy levels. Our calculations reveal that non-escaping regular orbits disappear for $E > 6$. Taking into consideration all the above-mentioned analysis we may conclude that in the configuration $(x,y)$ space exit channel 2 seems to be much more preferable at high energy levels with respect to exit channel 1.

\begin{figure*}[!t]
\centering
\resizebox{\hsize}{!}{\includegraphics{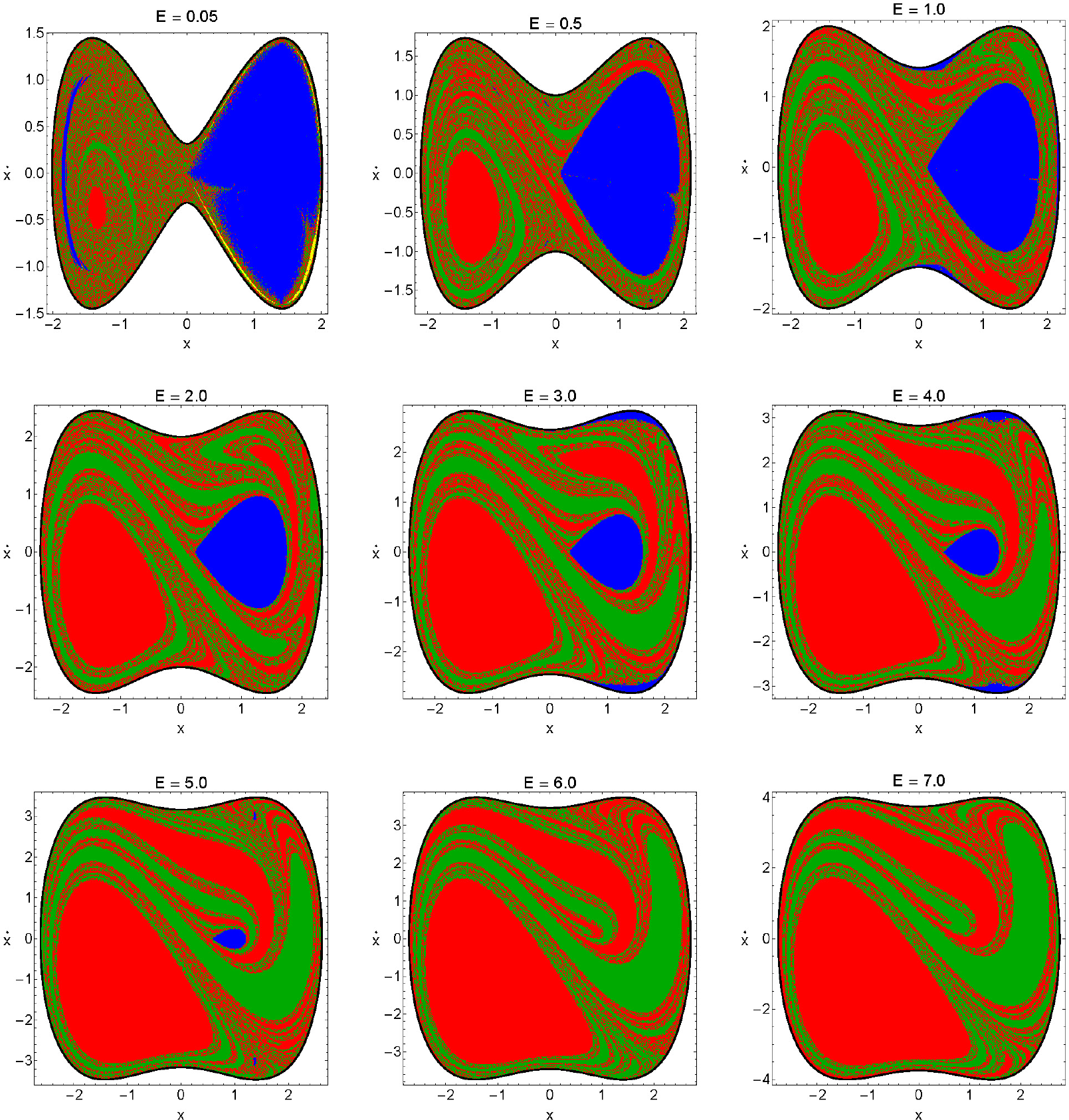}}
\caption{The orbital structure of the phase $(x,\dot{x})$ plane for several values of the energy $E$. The color code is he same as in Fig. \ref{xy}.}
\label{xpx}
\end{figure*}

\begin{figure*}[!t]
\centering
\resizebox{\hsize}{!}{\includegraphics{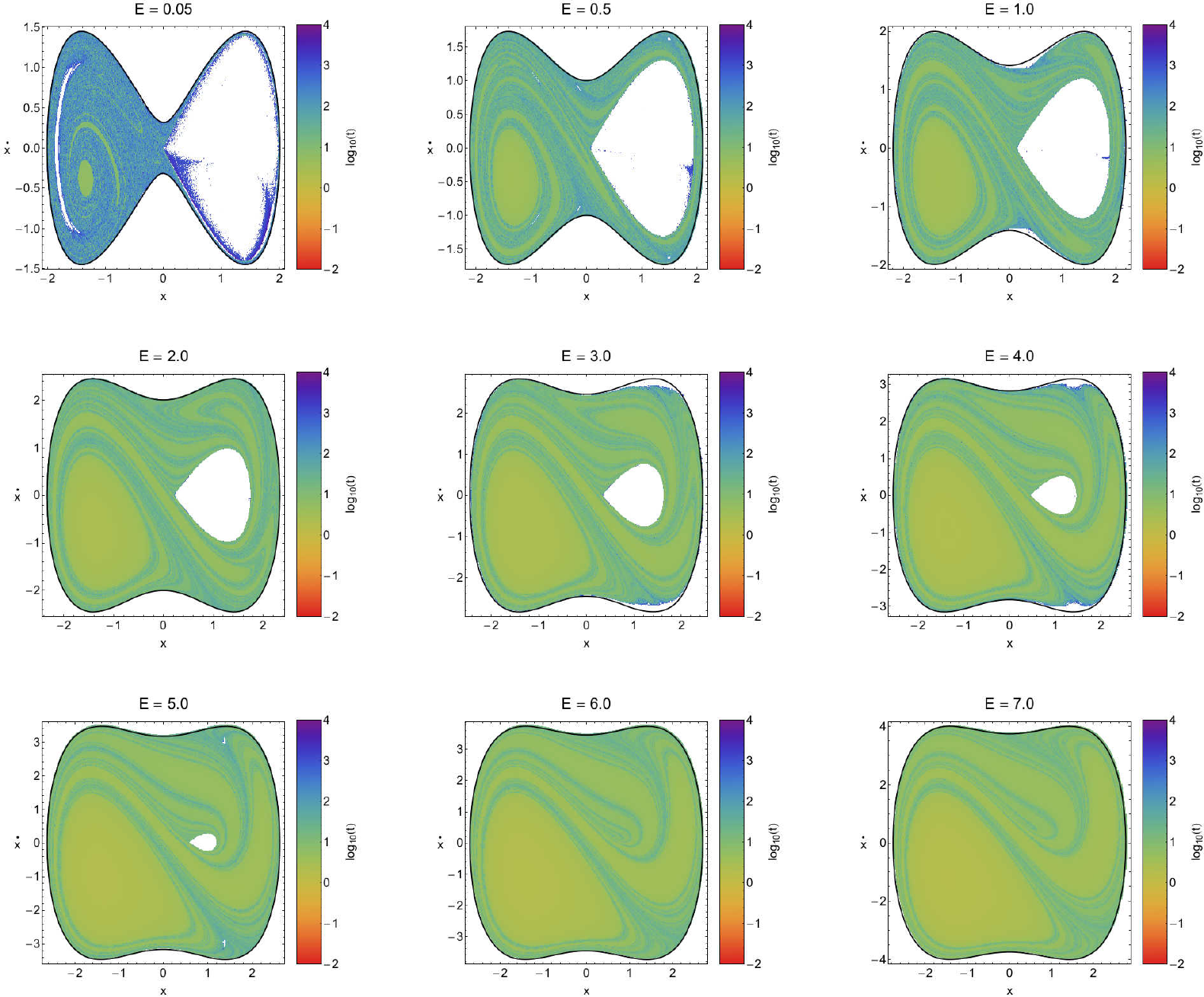}}
\caption{Distribution of the escape time $t_{\rm esc}$ of the orbits on the phase $(x,\dot{x})$ plane. The darker the color, the larger the escape time. Trapped chaotic and non-escaping regular orbits are shown in white.}
\label{xpxt}
\end{figure*}

\begin{figure*}[!t]
\centering
\resizebox{\hsize}{!}{\includegraphics{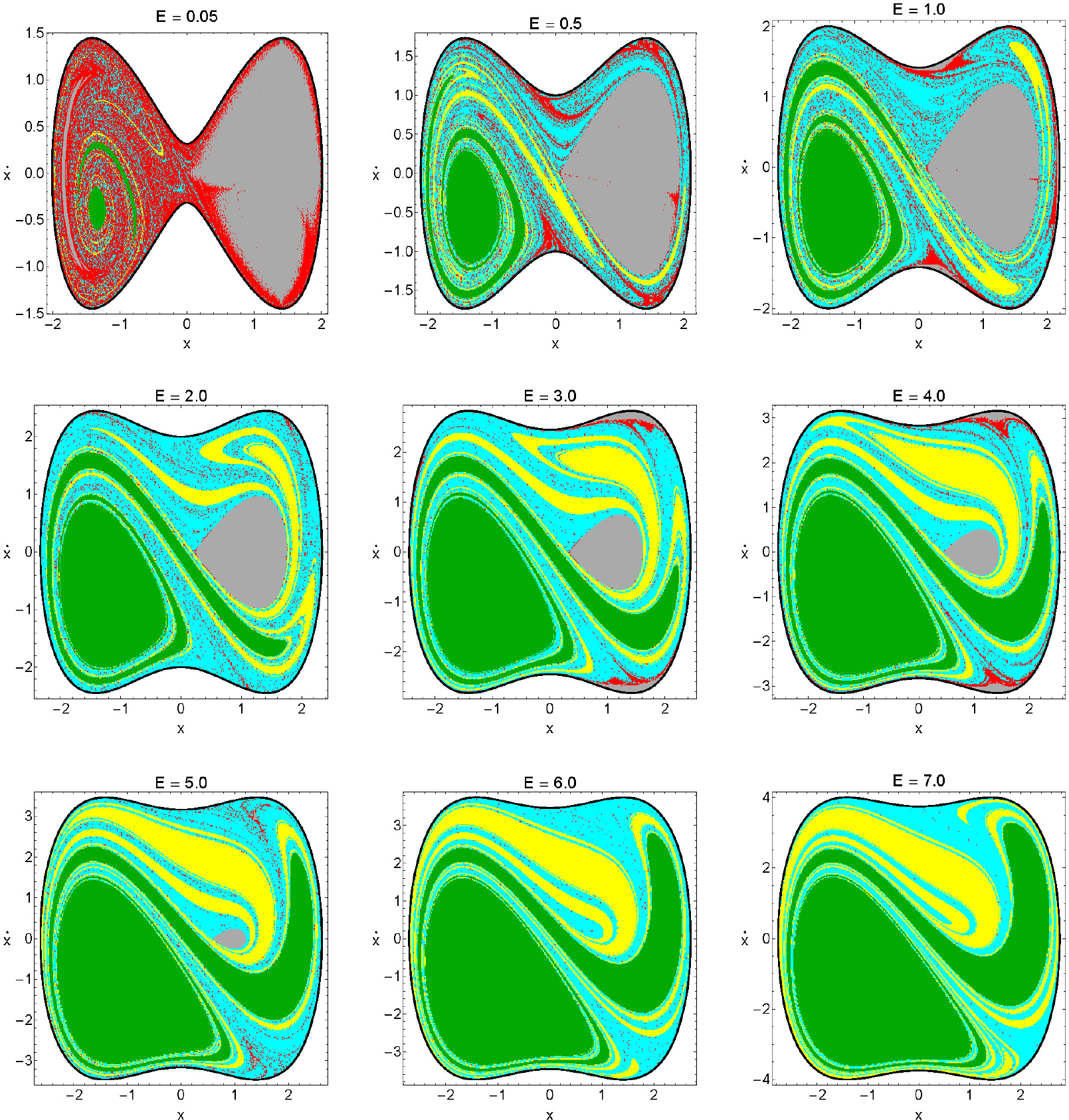}}
\caption{Color scale of the escape regions as a function of the number of intersections $N$ with the $y = 0$ axis upwards $(\dot{y} > 0)$. The color code is as follows: 0 intersections (green); 1 intersection (yellow); 2--10 intersections (cyan); $N > 10$ (red). The gray regions represent initial conditions of non-escaping regular and trapped chaotic orbits.}
\label{iters}
\end{figure*}

\begin{figure*}[!t]
\centering
\resizebox{\hsize}{!}{\includegraphics{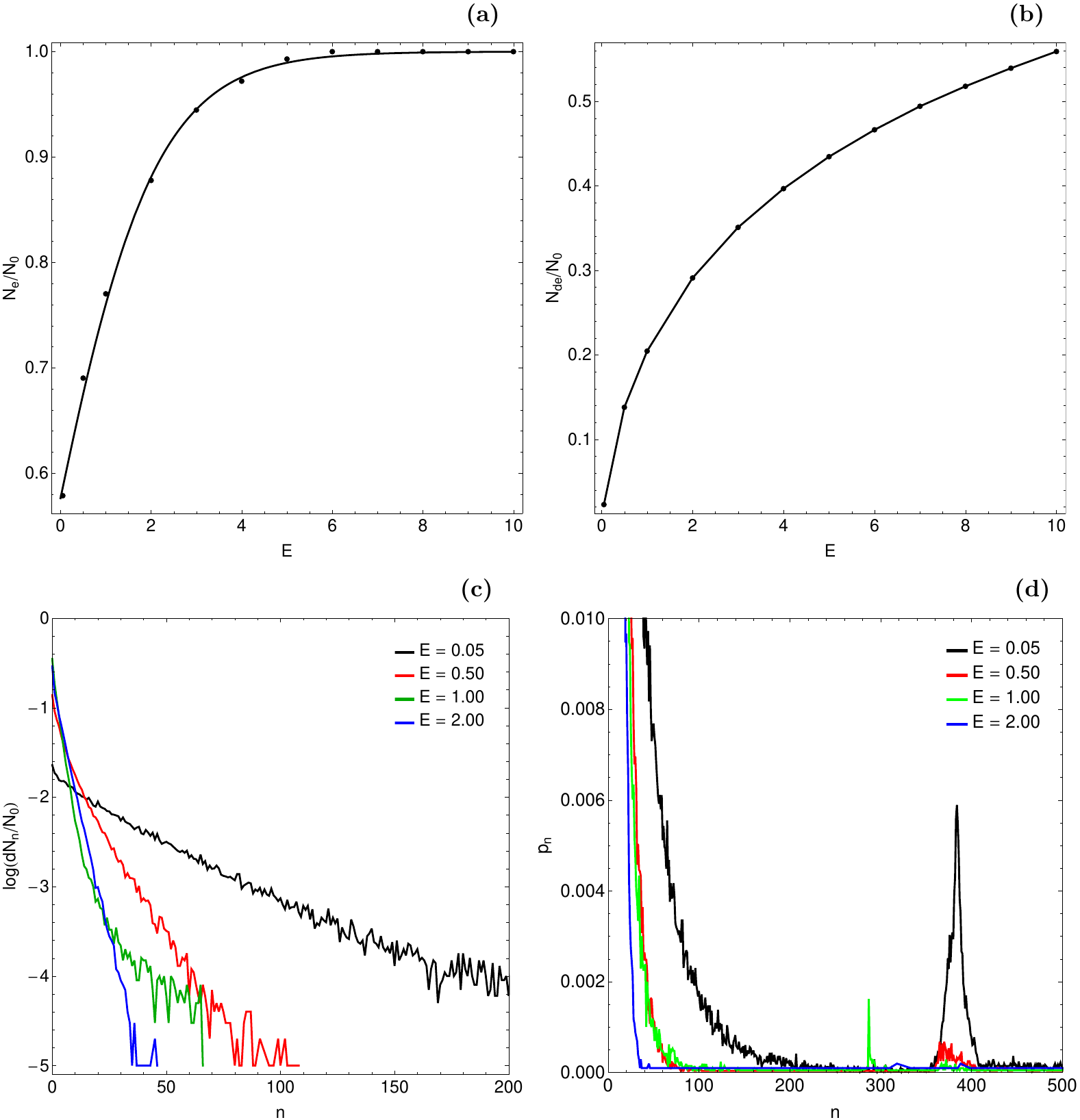}}
\caption{(a-upper left): Evolution of the proportion of escaping orbits $N_e/N_0$ as a function of the total orbital energy $E$, (b-upper right): Evolution of the proportion of directly escaping orbits $N_{de}/N_0$ as a function of the energy $E$, (c-lower left): Evolution of the logarithmic proportion $dN_n/N_0$ as a function of the number of the intersections $n$, for various values of the energy and (d-lower right): Evolution of the probability $p_n$ of escapes as a function of $n$ for several energy levels.}
\label{stats0}
\end{figure*}

\subsection{Results for the phase $(x,\dot{x})$ space}
\label{ss2}

For the phase $(x,\dot{x})$ space we consider orbits with initial conditions $(x_0, \dot{x_0})$ with $y_0 = 0$, while the initial value of $\dot{y}$ is obtained from the Hamiltonian (\ref{ham}). The orbital structure of the phase plane for the same set of values of the energy is shown in Fig. \ref{xpx}. A similar behavior to that discussed for the configuration $(x,y)$ plane can be seen. The outermost black solid line is the limiting curve which is defined as
\begin{equation}
f(x,\dot{x}) = \frac{1}{2}\dot{x}^2 + V_{D_5}(x, y = 0) = E.
\label{zvc}
\end{equation}
Here we must clarify that this $(x,\dot{x})$ phase plane is not a classical Poincar\'{e} Surface of Section (PSS), simply because escaping orbits in general, do not intersect the $y = 0$ axis after a certain time, thus preventing us from defying a recurrent time. A classical Poincar\'{e} surface of section exists only if orbits intersect an axis, like $y = 0$, at least once within a certain time interval. Nevertheless, in the case of escaping orbits we can still define local surfaces of section which help us to understand the orbital behavior of the dynamical system.

Looking at Fig. \ref{xpx} we see that the limiting curve is closed. This however does not mean that the test-particles cannot escape. It simply means that the escape channels are not visible in the phase space. Again, in the $(x,\dot{x})$ plane we can distinguish fractal regions where we cannot predict the particular channel of escape and regions occupied by escape basins. These basins are either broad well-defined regions, or elongated bands of complicated structure spiralling around the center. We see, once more, that for values of energy close to the escape energy there is a considerable amount of non-escaping orbits, occupying almost the entire right lobe, and the degree of fractalization of the phase plane is high. As we proceed to higher energy levels however, the rate of non-escaping regular orbits reduces, the phase plane becomes less and less fractal and well-defined basins of escape dominate. With a closer look in Fig. \ref{xpx} we can identify for $E = 0.05$ a thin filament around the main stability island which is composed of initial conditions of trapped chaotic orbits. The distribution of the escape time $t_{\rm esc}$ of orbits on the phase $(x,\dot{x})$ plane is shown in Fig. \ref{xpxt}. It is evident that orbits with initial conditions inside the exit basins escape from the system very quickly, or in other words, they possess extremely low escape periods. On the contrary, orbits with initial conditions located in the fractal basin boundaries need a considerable amount of time in order to escape.

Another interesting way of measuring the escape rate of an orbit is by counting how many intersection the orbit has with the axis $y = 0$ before it escapes. The regions in Fig. \ref{iters} are colored according to the number of intersections with the axis $y = 0$ upwards $(\dot{y} > 0)$. We observe, that orbits with initial conditions inside most of the basins escape directly without any intersection with the $y = 0$ axis. Furthermore, as the value of the energy increases, these green regions grow in relative size (proportion of the total area on the phase plane) and for high enough energy levels they occupy more than 50\% of the phase plane. We should also note, that orbits with initial conditions located at the fractal boundaries of the stability islands perform numerous intersections with the $y = 0$ axis before they eventually escape to infinity. On the other hand, orbits with initial conditions in the elongated spiral bands need only a couple of intersection until they escape. Similar types of plots showing the number of intersections can also be constructed for orbits with initial conditions in the configuration space.

It would be of particular interest to conduct a statistical analysis of the escape process in the case of the phase $(x,\dot{x})$ space. For this purpose, we shall follow the numerical approach used recently in \cite{CHLG12}. Our results are shown in Fig. \ref{stats0}(a-d) where curve fit approximation versus results from numerical integration is presented in the four panels. To begin with, Fig. \ref{stats0}a shows the proportion of escaping orbits $N_e/N_0$ as a function of the total orbital energy $E$. For large values of energy, $E > 6$, all the integrated orbits escape from the system. According to our numerical calculation, the evolution of the proportion of escaping orbits can be approximated by the formula
\begin{equation}
\frac{N_e}{N_0}(E) = 0.5\left[1 + \tanh\left(0.423667 E + 0.153542\right)\right].
\label{tr21}
\end{equation}
In Fig. \ref{stats0}b we present the evolution of the direct escaping orbits $N_{de}/N_0$ (by the term direct escaping orbits we refer to orbits that escape to infinity immediately without any intersection with the $x = 0$ axis) as a function of the energy $E$. We see, that the amount of direct escaping orbits grows rapidly with increasing $E$ and for high energy levels  they populate more than 50\% of the phase plane. The proportion of direct escapes can be given by the approximate formula
\begin{equation}
\frac{N_{de}}{N_0}(E) = 0.046 + 0.153 E - 0.019 E^2 + 0.001 E^3.
\label{tr22}
\end{equation}
Moreover, Fig. \ref{stats0}c depicts the logarithm of the proportion of escaping orbits $dN_n/N_0$, where $dN_n$ corresponds to the number of escaping orbits after the $n$th intersection with the $x = 0$ axis upwards $(\dot{x} > 0)$. It is seen, that the escape time of orbits decreases with increasing $n$. In particular, the escape rates are high for relatively small $n$, while they drop rapidly for larger $n$. Last but not least, we computed the probability of escape as a function of the number of intersections for various values of the energy. Specifically, the probability is defined as
\begin{equation}
p_n = \frac{d N_n}{N_n},
\label{tr3}
\end{equation}
where $N_n$ is the number of orbits that have not yet escaped before the $n$th intersection. The evolution of $p_n$ as a function of $n$ for various energy levels is given in Fig. \ref{stats0}d.

\begin{figure}[!t]
\begin{center}
\includegraphics[width=\hsize]{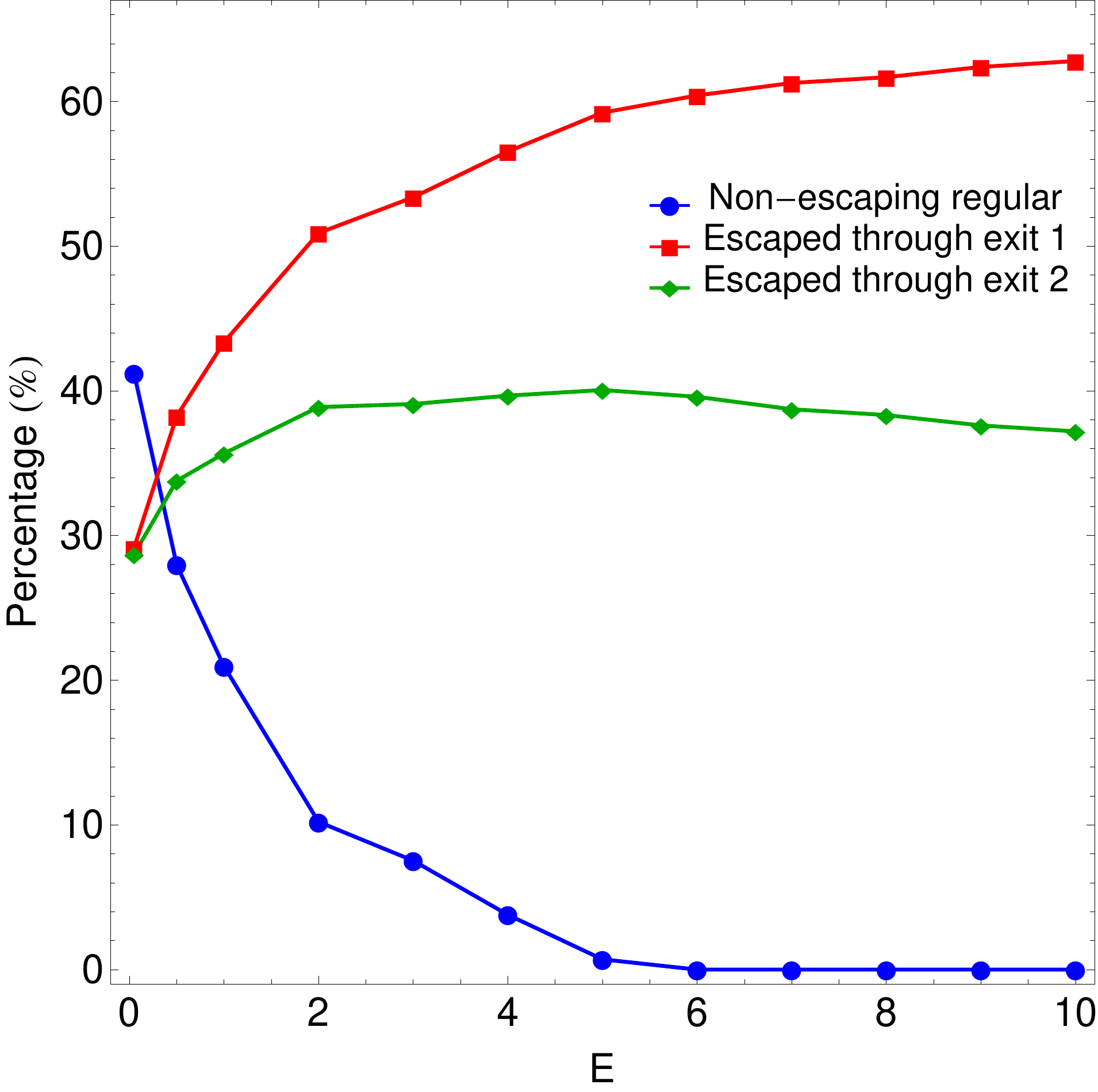}
\end{center}
\caption{Evolution of the percentages of escaping and non-escaping orbits with initial conditions on the phase $(x,\dot{x})$ space when varying the total orbital energy $E$.}
\label{p2}
\end{figure}

In Fig. \ref{p2}a we see the evolution of the percentages of escaping and non-escaping orbits on the phase $(x,\dot{x})$ plane when the value of the total orbital energy $E$ varies. Again the evolution of the percentage of trapped chaotic orbits was not included since the corresponding values are always very small (less than 1\%). One may observe that at low values of the energy non-escaping orbits is the most populated type of motion as they correspond to about 40\% of the phase space. With increasing energy however the rate of regular non-escaping orbits is reduced and for $E > 6$ it completely disappears. In Fig. \ref{p1} referring to the configuration space we seen that near the energy of escape the rates of escaping orbits coincide thus implying a high degree of fractalization. This is also true in the phase space although this time the rates of escaping orbits are slightly reduced with respect to the previous case. As the value of the energy increases the rates of escaping orbits start to diverge. Being more precise, the percentage of escaping orbits through exit channel 1 increases rapidly, while that of escaping orbits through exit channel 2 it exhibits almost a monotone behaviour around 38\% for $E > 2$. At the highest energy level studied $(E = 10)$ about 62\% of the phase space is occupied by initial conditions of orbits which escape through exit channel 1. By taking into account the results presented in Fig. \ref{p2} we may say that in the phase space and for high values of the energy, exit channel 1 seems to be more preferable.

\subsection{An overview analysis}
\label{over}

The color-coded grids in the configuration $(x,y)$ as well as the phase $(x,\dot{x})$ space provide information on the phase space mixing however, for only a fixed value of total orbital energy. H\'{e}non, back in the late 60s \cite{H69}, introduced a new type of plane which can provide information not only about stability and chaotic regions but also about areas of escaping and non-escaping orbits using the section $y = \dot{x} = 0$, $\dot{y} > 0$ (see also \cite{BBS08}). In other words, all the orbits of the test particles are launched from the $x$-axis with $x = x_0$, parallel to the $y$-axis $(y = 0)$. Consequently, in contrast to the previously discussed types of planes, only orbits with pericenters on the $x$-axis are included and therefore, the value of the energy $E$ can be used as an ordinate. In this way, we can monitor how the energy influences the overall orbital structure of our Hamiltonian system using a continuous spectrum of energy values rather than few discrete energy levels.

\begin{figure*}[!t]
\centering
\resizebox{\hsize}{!}{\includegraphics{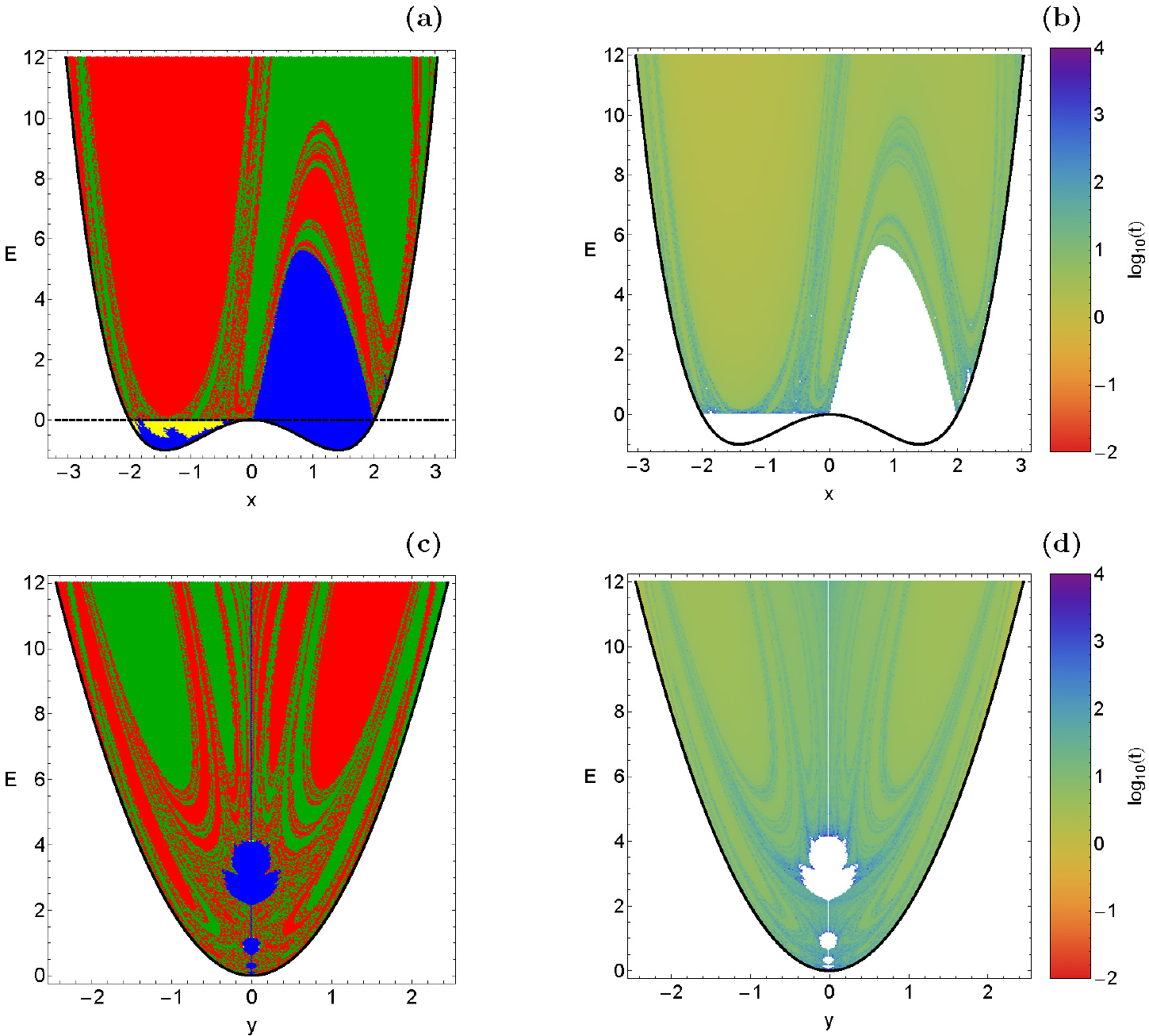}}
\caption{Orbital structure of the (a-upper left): $(x,E)$ plane and (c-lower left): $(y,E)$ plane. The color code is the same as in Fig. \ref{xy}. Panels (b) and (d): The distribution of the corresponding escape time of the orbits of the $(x,E)$ and $(y,E)$ plane, respectively.}
\label{xyEt}
\end{figure*}

\begin{figure*}[!t]
\centering
\resizebox{\hsize}{!}{\includegraphics{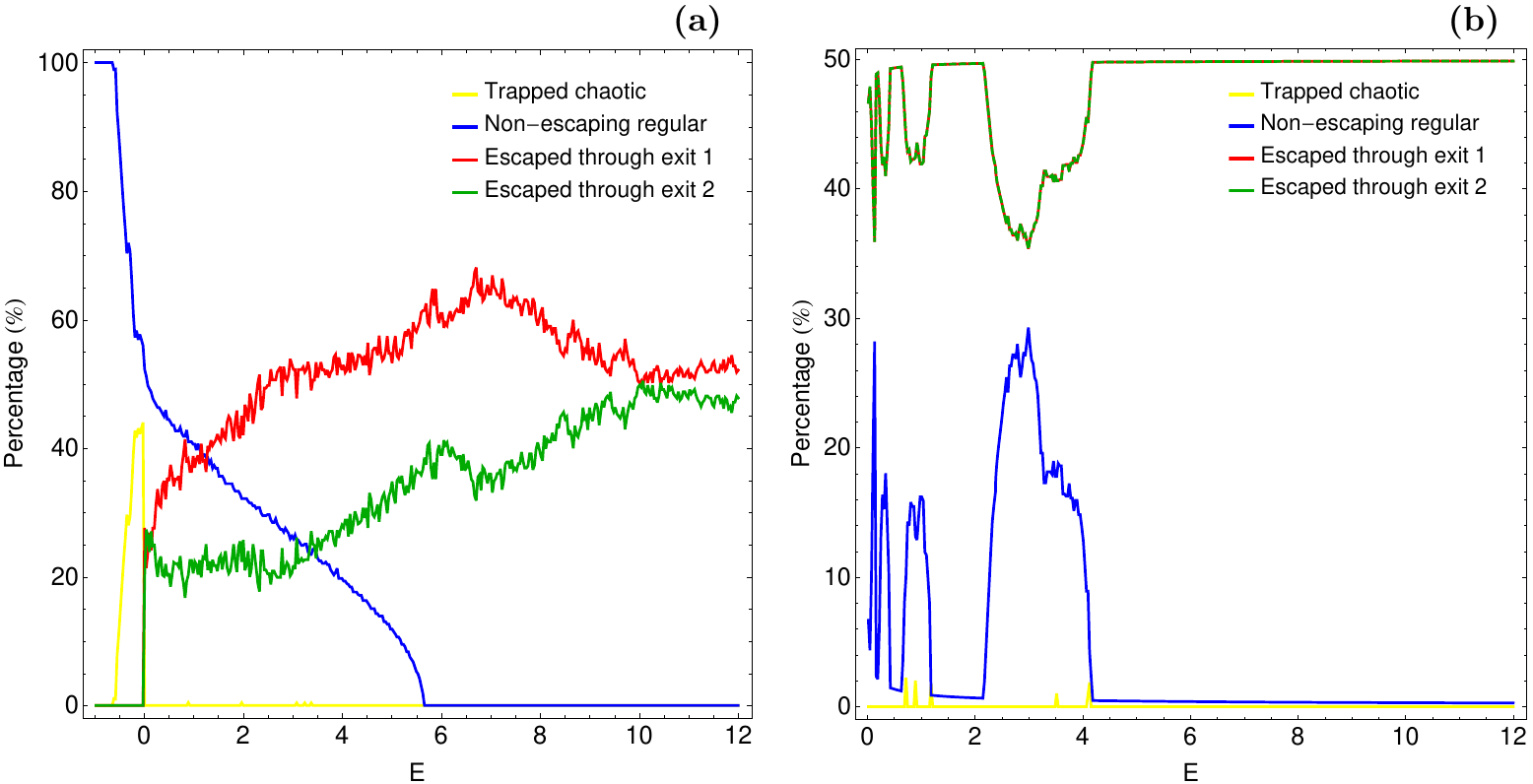}}
\caption{Evolution of the percentages of all types of orbits as a function of the total orbital energy $E$ in the (a-left): $(x,E)$ plane and (b-right): $(y,E)$ plane.}
\label{p3}
\end{figure*}

\begin{figure*}[!t]
\centering
\resizebox{\hsize}{!}{\includegraphics{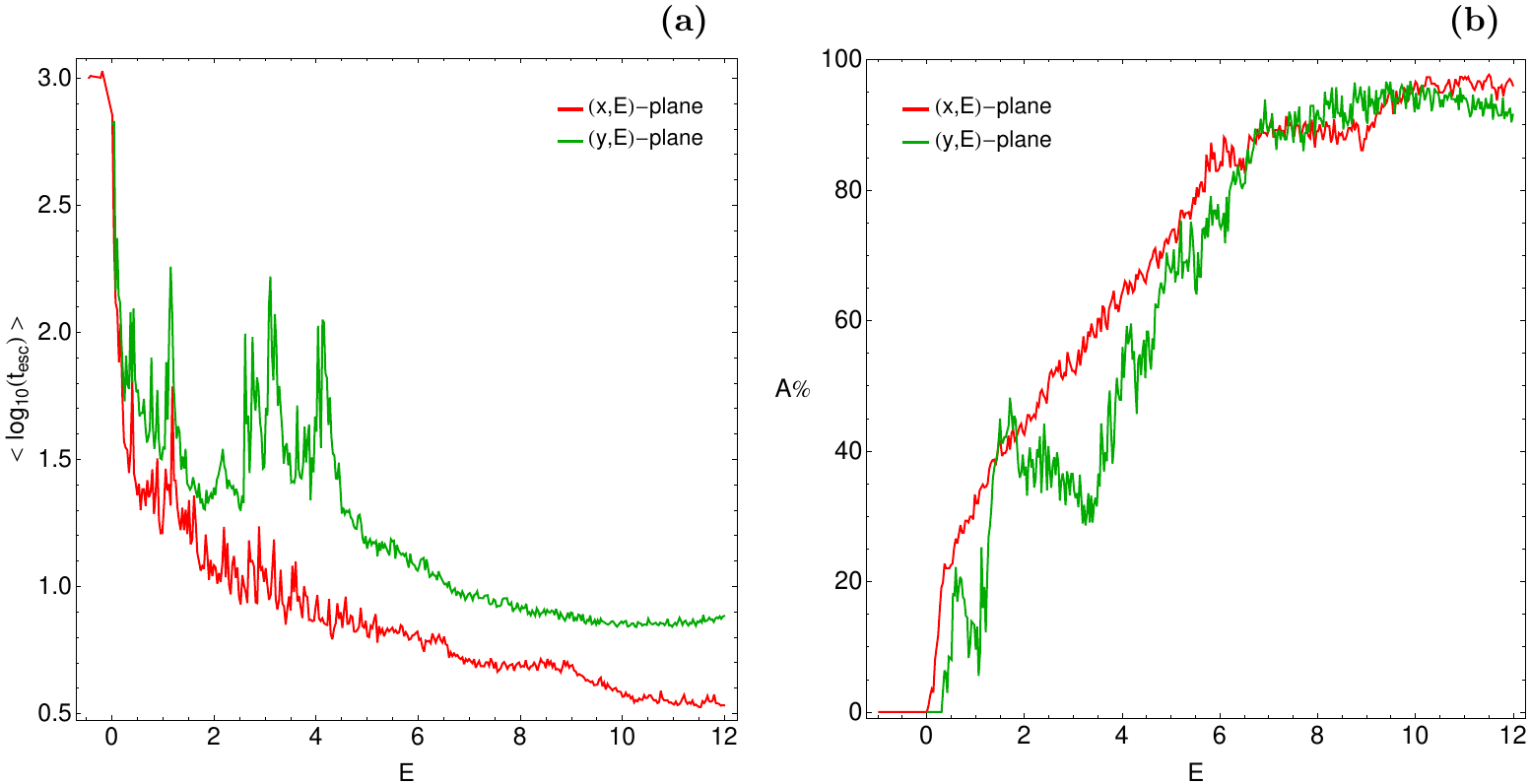}}
\caption{(a-left): The average escape time of orbits $< t_{\rm esc} >$ and (b-right): the percentage of the total area $A$ of the planes covered by the escape basins as a function of the total orbital energy $E$.}
\label{stats}
\end{figure*}

The orbital structure of the $(x,E)$ plane is presented in Fig. \ref{xyEt}a. The outermost black solid line is the limiting curve which is defined as
\begin{equation}
f_1(x,E) = V_{D_5}(x, y = 0) = E.
\label{zvc2}
\end{equation}
In this type of plane we can observe the two wells of $D_5$ potential. The horizontal black dashed line indicates the energy of escape $(E_{esc} = 0)$. It is seen that below the energy of escape there is a substantial amount of trapped chaotic orbits. All the initial conditions of these orbits are however confined only to the left will of the potential. For $E > 0$ trapped chaotic motion completely disappears. Well-defined basins of escape start to form for $E > 0$ and their extent grows rapidly with increasing energy, while fractal areas are observable only in the boundaries between the escape basins. For $x > 0$ a stability island is present however its size is reduced with increasing energy and at about $E = 5.5$ is vanishes. The value of the energy at which we observe the last indication of non-escaping regular motion is very close to the corresponding values reported earlier in both the configuration and the phase space. In Fig. \ref{xyEt}b we illustrate how the corresponding escape time of orbits are distributed on the $(x,E)$ plane.

In order to obtain a more complete view of the orbital structure of the dynamical system, we follow a similar numerical approach to that explained before but in this case we use the section $x = \dot{y} = 0$, $\dot{x} > 0$, considering orbits that are launched from the $y$-axis with $y = y_0$, parallel to the $x$-axis. This allow us to construct again a two-dimensional (2D) plane in which the $y$ coordinate of orbits is the abscissa, while the value of the energy $E$ is the ordinate. Fig \ref{xyEt}c shows the structure of the $(y,E)$ plane, while the distribution of the corresponding escape time of orbits is given in Fig. \ref{xyEt}d. The limiting curve this time is given by
\begin{equation}
f_2(y,E) = V_{D_5}(x = 0, y) = E.
\label{zvc3}
\end{equation}
It is interesting to note that the orbital structure of the $(y,E)$ plane is mirror symmetrical with respect to the $y$ axis $(x = 0)$.

Useful conclusions can be obtained by monitoring the evolution of the percentages of all types of orbits as a function of the value of the energy $E$. Fig. \ref{p3}(a-b) shows the diagrams corresponding to the $(x,E)$ and $(y,E)$ planes of Fig. \ref{xyEt}, respectively. In the $(x,E)$ plane for negative values of the energy regular motion dominates however the amount of non-escaping regular orbits significantly reduces and for about $E > 5.5$ it completely disappears. In the $(y,E)$ plane on the other hand, it displays fluctuations in the interval $E \in [0,4]$, while for larger values of the total orbital energy there is no indication of bounded regular motion. The evolution of the percentages of escaping orbits is also different in the two types of planes. In particular, in the $(x,E)$ plane the percentages of both escape channels coincide at about 25\%, just above the energy of escape, while for larger values of the energy the rates start to diverge. Our calculations suggest that in the energy range $E \in (0,12]$ the percentage of exit channel 1 is always larger than that of exit channel 2. In the $(y,E)$ plane things are quite different since both escape channels are always equiprobable due to the symmetry of the plane.

\begin{figure*}
\resizebox{\hsize}{!}{\includegraphics{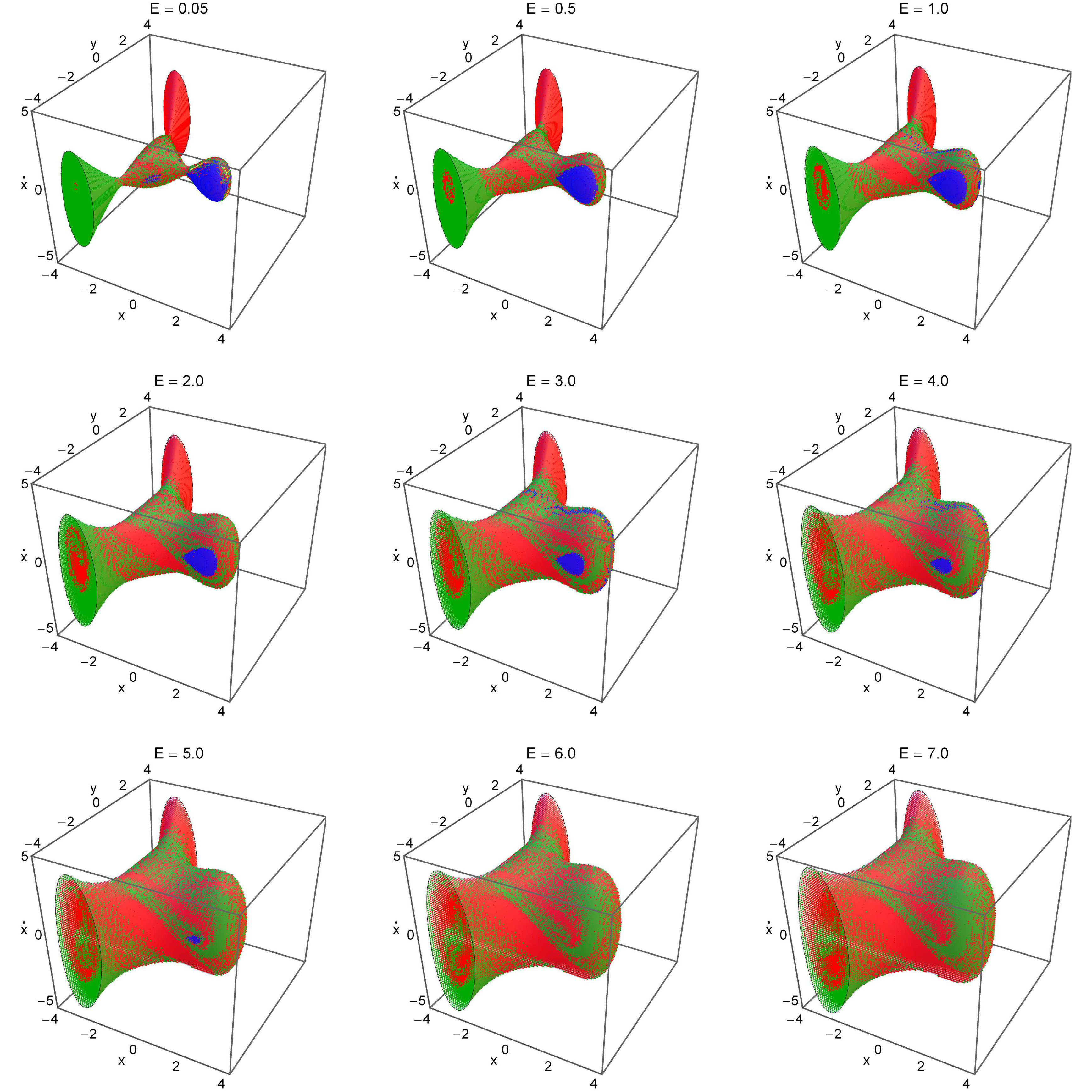}}
\caption{Orbital structure of three dimensional distributions of initial conditions of orbits in the 3D $(x,y,\dot{x})$ subspace for several values of the energy $E$. The color code is the same as in Fig. \ref{xy}.}
\label{3dgr}
\end{figure*}

Finally in Fig. \ref{stats}a we present the evolution of the average value of the escape time $< t_{\rm esc} >$ of orbits as a function of the total orbital energy for the $(x,E)$ and $(y,E)$ planes. We observe that for low energy levels, just above the critical energy of escape, the average escape period of orbits is more than 500 time units. However as the value of the energy increases the escape time of the orbits reduces rapidly, in general terms, tending asymptotically however to different values. If we want to justify the behaviour of the escape time we should take into account the geometry of the open ZVCs. In particular, as the total orbital energy increases the two symmetrical escape channels near the saddle points become more and more wide and therefore, the test particles need less and less time until they find one of the two openings (holes) in the ZVC and escape to infinity. This geometrical feature explains why for low values of the energy orbits consume large time periods wandering inside the open ZVC until they eventually locate one of the two exits and escape from the system. The evolution of the percentage of the total area $(A)$ on the $(x,E)$ and $(y,E)$ planes corresponding to basins of escape, as a function of the total orbital energy is given in Fig. \ref{stats}b. As expected, for low values of the energy the degree of fractality on both types of planes is high. However, as the energy increases the rate of fractal domains reduces and the percentage of domains covered by basins of escape starts to grow rapidly. Eventually, at relatively high energy levels $(E > 7)$ the fractal domains are significantly confined and therefore the well formed basins of escape occupy more than 90\% of both types of planes.

\subsection{Three dimensional distributions of initial conditions of orbits}
\label{3d}

In all previous subsections we investigated the escape dynamics of orbits using two dimensional grids of initial conditions in several types of planes (or in other words, in several 2D subspaces of the whole 4D phase space). Unfortunately the results presented in these subsections are different between each other, both from a qualitative and from a quantitative point of view. This negatively affect the conclusions that have been drawn, which are evidently ``subspace- dependent". Therefore, we decided to expand our numerical exploration using three dimensional distributions of initial conditions of orbits. Being more precise, for a particular value of the energy we define inside the corresponding zero velocity surface uniform grids of $100 \times 100 \times 100$ initial conditions $(x_0,y_0,\dot{x_0})$, while the initial value of $\dot{y} > 0$ is always obtained from the energy integral of motion (\ref{ham}).

In Fig. \ref{3dgr} we present the orbital structure of the three dimensional distributions of initial conditions of orbits in the $(x,y,\dot{x})$ subspace for the same set of values of the energy. The color code is the same as in Fig. \ref{xy}. It is seen that in this case the grid of the initial conditions of the orbits is in fact a three dimensional solid and therefore only its outer surface is visible. However, a tomographic-style approach can be used in order to penetrate and examine the interior region of the solid (e.g., \cite{Z14a}). In the previous subsections \ref{ss1} and \ref{ss2} we classified initial conditions of orbits in 2D subspaces (see Figs. \ref{xy} and \ref{xpx}) which can be considered as horizontal and vertical tomographic slices of the corresponding three dimensional solids shown in Fig. \ref{3dgr}. Thus we may say that we managed to obtain a more general and spherical view of the escape dynamics of the $D_5$ potential. Here we would like to point out that to our knowledge this is the first time that a three dimensional distribution of initial conditions of orbits is used for classifying orbits in open Hamiltonian systems.

\begin{figure}
\resizebox{\hsize}{!}{\includegraphics{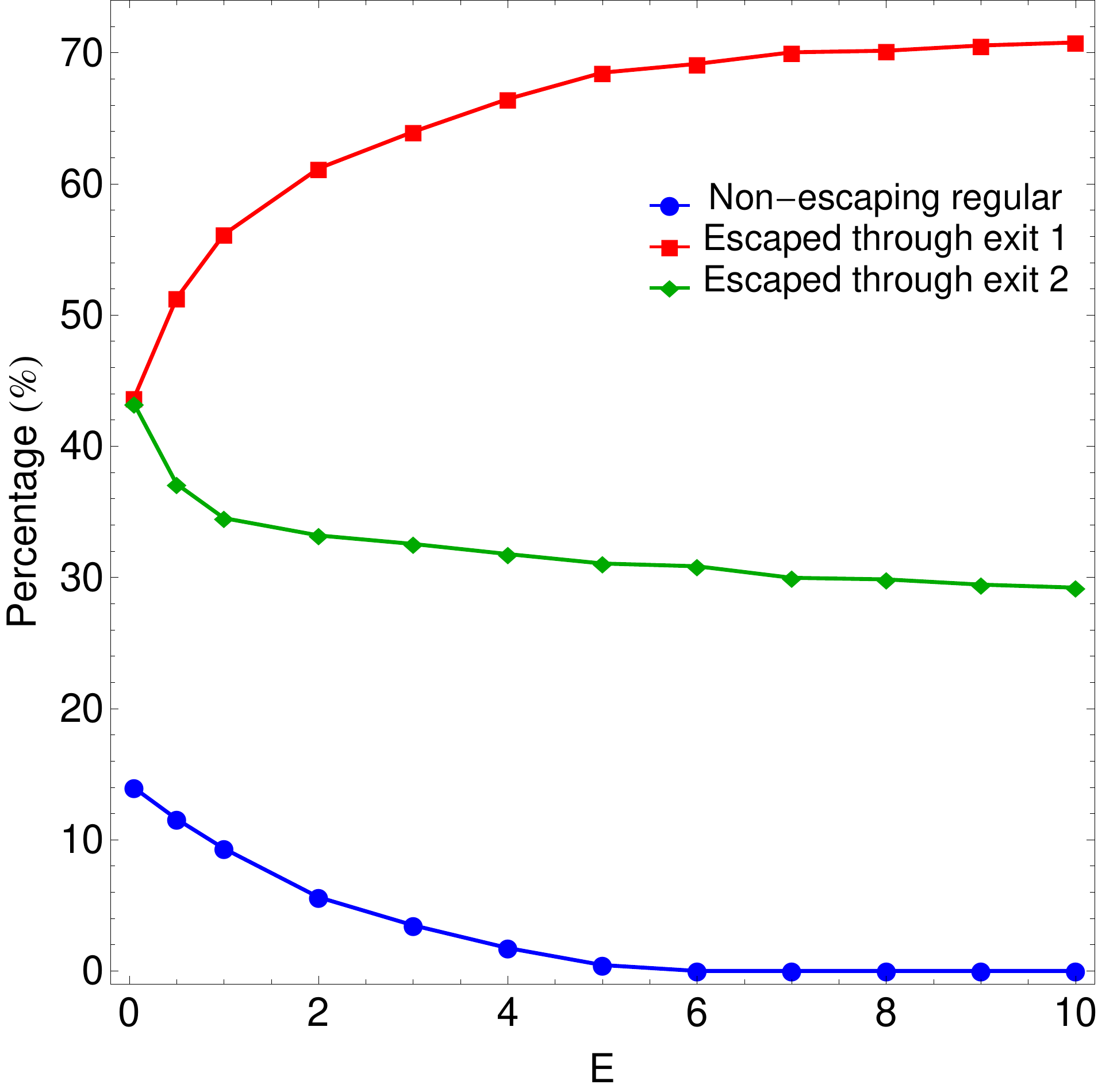}}
\caption{Evolution of the percentages of escaping and non-escaping orbits with initial conditions inside the three dimensional grids, as a function of the total orbital energy $E$.}
\label{p3d}
\end{figure}

Using the above-mentioned technique regarding the three dimensional distribution of initial conditions of orbits we can monitor how the percentages of the different types of orbits with initial conditions inside the 3D $(x,y,\dot{x})$ subspace evolve as a function of the total orbital energy $E$. For obtaining a more complete view we integrated more 3D grids (not shown here) for additional higher energy levels $(E > 7)$. Our results are presented in Fig. \ref{p3d}. Once more the evolution of the percentage of trapped chaotic orbits was not included since the corresponding values are always extremely small (less than 0.5\%). Escaping orbits share about 86\% of the 3D subspace for $E = 0.05$, that is an energy level just above the energy of escape. With increasing energy however, the rates of escaping orbits start to diverge. In particular, our calculations indicate that the percentage of orbits that escape through exit 1 increases, while the rate of orbits that escape through exit channel 2 decreases. The amount of non-escaping regular orbits also decreases as we proceed to higher energy levels and for $E > 6$ they completely disappear. Our analysis suggests that for $E > 7$ the rates of escaping orbits seem to saturate, around 70\% for channel 1 and around 30\% for channel 2. Thus we may conclude that in the 3D $(x,y,\dot{x})$ subspace and especially for relatively high energy levels the probability of an orbit escaping through exit channel 1 is double with respect to exit channel 2.

\section{Discussion}
\label{disc}

The aim of this work was to numerically investigate the escape dynamics in the $D_5$ potential which is a characteristic example of a two-dimensional multiwell potential. For this purpose we investigated the orbital structure in many types of two-dimensional planes and for several values of the total orbital energy $E$. We also proceeded one step further by classifying initial conditions of orbits in a three-dimensional (3D) subspace of the whole four-dimensional (4D) phase space. We managed to distinguish between escaping and non-escaping orbits and we also located the basins of escape leading to different exit channels, finding correlations with the corresponding escape time of the orbits. Among the escaping orbits we separated between those escaping fast or late from the system. Our extensive numerical calculations strongly suggest that the overall escape process in the $D_5$ potential is very dependent on the value of the total orbital energy.

We also performed a statistical analysis, relating the proportion of escaping and directly escaping orbits with the value of the energy. In the same vein, the evolution of the proportion of escaping orbits and the corresponding probability, as functions of the $n$-th intersection with the $y = 0$ axis upwards was also presented. As far as we know, this is the first time that the escape process in a two-dimensional multiwell potential is explored through orbit classification in such a detailed and systematic way and this is exactly the novelty and the contribution of the current work.

The multivariate Newton-Raphson root method was successfully used in order to obtain the basins of attraction corresponding to the five equilibrium points. A highly complicated structure has been revealed on the configuration $(x,y)$ plane. Well-defined basins of attractions with fractal boundaries have found to dominate the configuration plane.

For the numerical integration of the sets of the initial conditions of orbits in each type of plane, we needed between about 3 minutes and 20 hours of CPU time on a Quad-Core i7 2.4 GHz PC, depending of course on the escape rates of orbits in each case. For each initial condition the maximum time of the numerical integration was set to be equal to $10^4$ time units however, when a test particle escaped the numerical integration was effectively ended and proceeded to the next available initial condition.

We hope that the present thorough analysis and the corresponding numerical results of the $D_5$ potential to be useful in this active field of nonlinear Hamiltonian systems by shedding some light to the complicated escape mechanism of orbits. Since our results are encouraging, it is in our future plans to investigate the escape properties of orbits in the $D_7$ potential which has three wells.

\section*{Acknowledgments}

I would like to express my warmest thanks to the four anonymous referees for the careful reading of the manuscript and for all the apt suggestions and comments which allowed us to improve both the quality and the clarity of the paper.

\section*{Compliance with Ethical Standards}

\begin{itemize}
  \item Funding: The author states that he has not received any research grants.
  \item Conflict of interest: The author declares that he has no conflict of interest.
\end{itemize}

\section*{Appendix: Derivation of the multivariate Newton-Raphson iterative scheme}
\label{apex}

The multivariate Newton-Raphson method iterates over the recursive formula
\begin{equation}
{\bf{x}}_{n+1} = {\bf{x}}_{n} - J^{-1}f({\bf{x}}_{n}),
\label{sch}
\end{equation}
where $J^{-1}$ is the inverse Jacobian matrix of $f({\bf{x}}_{n})$, while ${\bf{x}}_{n}$ stands for the vector $x$ at the $n$-th iteration. In our case the system of the equations is $V_x = 0$ and $V_y = 0$ and therefore the Jacobian matrix reads
\begin{equation}
J =
\begin{bmatrix}
V_{xx} & V_{xy} \\
V_{yx} & V_{yy}
\end{bmatrix}.
\label{jac}
\end{equation}
The inverse Jacobian is
\begin{equation}
J^{-1} = \frac{1}{{\rm{det}}(J)}
\begin{bmatrix}
V_{yy} & -V_{xy} \\
-V_{yx} & V_{xx}
\end{bmatrix},
\label{ijac}
\end{equation}
where ${\rm{det}}(J) = V_{yy} V_{xx} - V_{xy}^2$.

Inserting the expression of the inverse Jacobian into the iterative formula (\ref{sch}) we get
\begin{eqnarray}
\begin{bmatrix}
x \\
y
\end{bmatrix}
_{n+1} &=&
\begin{bmatrix}
x \\
y
\end{bmatrix}
_{n} - \frac{1}{V_{yy} V_{xx} - V_{xy}^2}
\begin{bmatrix}
V_{yy} & -V_{xy} \\
-V_{yx} & V_{xx}
\end{bmatrix}
\begin{bmatrix}
V_x \\
V_y
\end{bmatrix}
_{(x_n,y_n)}
\nonumber\\
&=&
\begin{bmatrix}
x \\
y
\end{bmatrix}
_{n} - \frac{1}{V_{yy} V_{xx} - V_{xy}^2}
\begin{bmatrix}
V_{yy}V_x - V_{xy}V_y \\
-V_{yx}V_x + V_{xx}V_y
\end{bmatrix}
_{(x_n,y_n)}.
\label{sch2}
\end{eqnarray}

Decomposing formula (\ref{sch2}) into $x$ and $y$ we obtain the iterative formulae for each coordinate
\begin{eqnarray}
x_{n+1} &=& x_n - \left( \frac{V_x V_{yy} - V_y V_{xy}}{V_{yy} V_{xx} - V^2_{xy}} \right)_{(x_n,y_n)}, \nonumber\\
y_{n+1} &=& y_n + \left( \frac{V_x V_{yx} - V_y V_{xx}}{V_{yy} V_{xx} - V^2_{xy}} \right)_{(x_n,y_n)}.
\end{eqnarray}

\end{document}